\makeatletter
\let\latexkernel@label\label
\makeatother

\documentclass[%
reprint,
superscriptaddress,
amsmath,amssymb,
aps,
showkeys,
pra,
]{revtex4-2}

\makeatletter
\let\label\latexkernel@label
\makeatother

\usepackage{varwidth}
\usepackage{amsmath}
\usepackage{amsthm}
\usepackage{amssymb}
\usepackage{graphicx}
\usepackage{hyperref}
\usepackage{subcaption}
\usepackage{orcidlink} 
\usepackage{comment}
\usepackage{multirow}
\usepackage{physics}
\usepackage{bbm}
\usepackage{hhline}

\makeatletter

\theoremstyle{plain}
\newtheorem{thm}{\protect\theoremname}
\theoremstyle{definition}

\theoremstyle{plain}
\newtheorem*{lem*}{\protect\lemmaname}

\let\Tr\relax

\DeclareMathOperator{\Tr}{Tr}

\DeclareMathOperator{\argmin}{argmin}
\newcommand{\1}{\leavevmode{\mathrm{1\ifmmode\mkern  -4.8mu\else\kern -.3em\fi I}}}

\makeatother

\providecommand{\definitionname}{Definition}
\providecommand{\lemmaname}{Lemma}
\providecommand{\theoremname}{Theorem}

\newcommand{\ignore}[1]{}
\newcommand{\hi}{\mathcal{H}}
\newcommand{\pur}{\mathcal{P}}
\newcommand{\exv}{\mathbb{E}}
\newcommand{\id}{\mathbbm{1}}
\newcommand{\Id}{\mathbbm{1}}

\newcommand{\cnot}{\operatorname{CNOT}}
\newcommand{\ot}{\otimes}
\newcommand{\ba}{\begin{eqnarray}}
\newcommand{\ea}{\end{eqnarray}}
\renewcommand{\var}{{\rm Var}}
\newtheorem{lemma}{Lemma}

\renewcommand{\selectlanguage}[1]{}

\begin{document}

\title{Experimental demonstration of non-local magic in a superconducting quantum processor}% Force line breaks with \\

\date{\today}% It is always \today, today,

\title{Experimental demonstration of non-local magic in a superconducting quantum processor }

% Use letters for affiliations, numbers to show equal authorship (if applicable) and to indicate the corresponding author

\author{Halima Giovanna Ahmad}
\thanks{These authors equally contributed to this work.}
\affiliation{%
	Dipartimento di Fisica “Ettore Pancini”, Università degli Studi di Napoli “Federico II”, Complesso Universitario di Monte Sant'Angelo, Via Cinthia, 21, Napoli, 80126, Italy
}%
\author{Gianluca Esposito${}^{*}$}
\affiliation{%
	Mathematical and Physical Sciences for Advanced Material and Technologies, Scuola Superiore Meridionale, Via Mezzocannone, 4, Napoli, 80134, Italy.
}%
\affiliation{Istituto Nazionale di Fisica Nucleare (INFN), Sezione di Napoli, Complesso Universitario di Monte Sant'Angelo, Via Cinthia, 21, Napoli, 80126, Italy}%Lines break automatically or can be forced with \\
\author{Viviana Stasino}
\affiliation{%
	Dipartimento di Fisica “Ettore Pancini”, Università degli Studi di Napoli “Federico II”, Complesso Universitario di Monte Sant'Angelo, Via Cinthia, 21, Napoli, 80126, Italy
}%
\author{Jovan Odavi\'c}
\affiliation{Dipartimento di Fisica “Ettore Pancini”, Università degli Studi di Napoli “Federico II”, Complesso Universitario di Monte Sant'Angelo, Via Cinthia, 21, Napoli, 80126, Italy
}%
\affiliation{Istituto Nazionale di Fisica Nucleare (INFN), Sezione di Napoli, Complesso Universitario di Monte Sant'Angelo, Via Cinthia, 21, Napoli, 80126, Italy}%Lines break automatically or can be forced with \\
\author{Carlo Cosenza}
\affiliation{%
	Dipartimento di Fisica “Ettore Pancini”, Università degli Studi di Napoli “Federico II”, Complesso Universitario di Monte Sant'Angelo, Via Cinthia, 21, Napoli, 80126, Italy
}%
\author{Alessandro Sarno}
\affiliation{%
	Dipartimento di Fisica “Ettore Pancini”, Università degli Studi di Napoli “Federico II”, Complesso Universitario di Monte Sant'Angelo, Via Cinthia, 21, Napoli, 80126, Italy
}%
\author{Pasquale Mastrovito}
\affiliation{%
	Dipartimento di Fisica “Ettore Pancini”, Università degli Studi di Napoli “Federico II”, Complesso Universitario di Monte Sant'Angelo, Via Cinthia, 21, Napoli, 80126, Italy
}%
\author{Michele Viscardi}
\affiliation{%
	Dipartimento di Fisica “Ettore Pancini”, Università degli Studi di Napoli “Federico II”, Complesso Universitario di Monte Sant'Angelo, Via Cinthia, 21, Napoli, 80126, Italy
}%
\affiliation{Istituto Nazionale di Fisica Nucleare (INFN), Sezione di Napoli, Complesso Universitario di Monte Sant'Angelo, Via Cinthia, 21, Napoli, 80126, Italy}%Lines break automatically or can be forced with \\
\author{Stefano Cusumano}
\affiliation{%
	Dipartimento di Fisica “Ettore Pancini”, Università degli Studi di Napoli “Federico II”, Complesso Universitario di Monte Sant'Angelo, Via Cinthia, 21, Napoli, 80126, Italy
}%
\affiliation{Istituto Nazionale di Fisica Nucleare (INFN), Sezione di Napoli, Complesso Universitario di Monte Sant'Angelo, Via Cinthia, 21, Napoli, 80126, Italy}%Lines break automatically or can be forced with \\
\author{Francesco Tafuri}%
\email{francesco.tafuri@unina.it}
\affiliation{%
	Dipartimento di Fisica “Ettore Pancini”, Università degli Studi di Napoli “Federico II”, Complesso Universitario di Monte Sant'Angelo, Via Cinthia, 21, Napoli, 80126, Italy
}%
\affiliation{Physics Department, XYZ University.}
\author{Davide Massarotti}
\affiliation{%
	Dipartimento di Ingegneria Elettrica e delle Tecnologie per l'Informazione, Via Claudio, Napoli, 80125, Italy
}%
\author{Alioscia Hamma}%
\email{alioscia.hamma@unina.it}
\affiliation{%
	Dipartimento di Fisica “Ettore Pancini”, Università degli Studi di Napoli “Federico II”, Complesso Universitario di Monte Sant'Angelo, Via Cinthia, 21, Napoli, 80126, Italy
}%
\affiliation{Mathematical and Physical Sciences for Advanced Material and Technologies, Scuola Superiore Meridionale, Via Mezzocannone, 4, Napoli, 80134, Italy.}
\affiliation{Istituto Nazionale di Fisica Nucleare (INFN), Sezione di Napoli, Complesso Universitario di Monte Sant'Angelo, Via Cinthia, 21, Napoli, 80126, Italy}

% At least three keywords are required at submission. Please provide three to five keywords, separated by the pipe symbol.

\begin{abstract}
Non-local magic is the non-stabilizerness that no local unitary operation can erase. It captures the joint action of entanglement and magic underlying quantum advantage, and it has never been measured on quantum hardware. Here we report its first experimental demonstration, on a superconducting quantum processing unit, through two independent routes: an optimal local-erasure protocol and a direct, state-agnostic measurement of subsystem purity. The two agree with each other and with theory. Exploiting direct access to the device, we construct a noise model with no free parameters that identifies readout error and a depolarizing controlled-Z channel as the dominant mechanisms, and we show that local and non-local magic can be addressed separately, erasing local magic in situ while preserving the non-local part. Non-local magic provides a hardware benchmark beyond standard gate-fidelity protocols and points toward more reliable pre-fault-tolerant devices. The same tools underlie a purity-estimation protocol with exponential speedup and the decoding of Hawking radiation in a black-hole toy model.
\end{abstract}

\maketitle

\section*{Introduction}

An essential objective in quantum information science is to identify the physical resources that lead to quantum computational speedups over classical algorithms~~\cite{feynman1982simulating,shor1997polynomial, harrow2009quantum,lloyd1996universal, grover1997quantum}. 
Entanglement is necessary for universal fault-tolerant quantum computation, though not sufficient for quantum advantage. A common paradigm for fault-tolerant operations uses the stabilizer formalism~~\cite{gottesman1998heisenberg,aaronsonimproved}.
 In this framework, the additional resources needed for quantum advantage are non-stabilizer states and operations, namely those resources that go beyond Clifford operations, also colloquially known as \textit{magic}~\cite{Bravyi_Kitaev_2005, Veitch_HamedMousavian_Gottesman_Emerson_2014, Howard_Campbell_2017}. 
 Stabilizer operations can generate extensive entanglement but no magic, whereas local operations can inject extensive magic but no entanglement. In both cases the resulting states are efficiently simulable on a classical computer, so quantum complex behavior and quantum advantage emerge only from the interplay of the two resources~\cite{iannotti2025coventmag, tirritoflatness,  Szombathy_Valli_Moca_Farkas_Zarand_2025,10.21468/SciPostPhys.9.6.087,piroli_2025,Odavic_Viscardi_Hamma_2025}. 

\textit{Non-local magic} $M_2^{\rm NL}$~\cite{Cao_Cheng_Hamma_Leone_Munizzi_Oliviero_2024} is a recent notion introduced to capture the interplay between entanglement and non-stabilizerness. It quantifies the fraction of non-stabilizer resources that cannot be removed by local unitary operations and operationally represents the amount of non-stabilizer resources that cannot be distilled using local operations. This construct is useful as it constitutes a form of bound resource, that is, a kind of resource that cannot be distilled by a suitable class of operations. For instance, this happens if the non-stabilizer resource is injected locally but then scrambled around, and cannot be extracted again by local operations. In the context of Measurement-Based computation, a related notion is that of invested and potential magic resources \cite{4yyv-hggz}. 

Even in the noisy intermediate-scale quantum (NISQ)~\cite{Preskill2018} era, local and non-local stabilizer resources  are both believed to be crucial for universal quantum  computation~\cite{4yyv-hggz}: in quantum algorithms that rely on the preparation of complex quantum states exhibiting $M_2^{\rm NL}$~\cite{Odavic_Haug_Torre_Hamma_Franchini_Giampaolo_2023}, or in magic state distillation processes in fault-tolerant quantum computing, where stabilizer codes are used to purify non-stabilizer states and achieve fault-tolerant non-Clifford operations~\cite{Wills2025,SalesRodriguez2025}. Notably, non-local stabilizer resources can also be a hindrance to some genuine quantum behavior, for instance, the violation of Bell's inequalities~\cite{cusumano2025nonstabilizernessviolationschshinequalities}. Moreover, $M_2^{\rm NL}$ plays a role in the field of Anti-de Sitter/Conformal Field Theory (AdS/CFT) correspondence, representing the holographic counterpart of gravitational back-reaction~\cite{Cao_Cheng_Hamma_Leone_Munizzi_Oliviero_2024} and provides a connection between quantum advantage, error-correcting codes and holography~\cite{PastawskiYoshidaHarlowPreskill2015}.

We access non-local magic through two complementary protocols, both built on the Stabilizer R\'enyi Entropy (SRE) as the unique computable measure of non-stabilizerness for pure states~\cite{Leone_Oliviero_Hamma_2022,Leone_Bittel_2024,Oliviero_Leone_Hamma_Lloyd_2022}, estimated here with the randomized-measurement toolbox~\cite{Elben_Flammia_Huang_Kueng_Preskill_Vermersch_Zoller_2022}. The first protocol is state-dependent: it quantifies non-local magic as the SRE that survives an optimal local erasure of the state's removable magic. The second is state-agnostic, rests on theoretical tools developed here, and extracts non-local magic directly from subsystem purity. Beyond corroborating each other, the two carry a metrological dividend: erasing local magic is the elementary cleansing primitive behind an algorithm that estimates subsystem purity, and hence entanglement, with an exponential speedup~\cite{Leone_Oliviero_Esposito_Hamma_2024}. Our experiments use two qubits, but both protocols, the state-agnostic one included, are in principle scalable.

A second aim of this work is to establish non-local magic as a probe for characterizing quantum hardware. Resolving it on a real device forces a quantitative confrontation with noise, and that confrontation is itself informative: the two dominant error channels map onto the two kinds of magic. Readout error acts as a local perturbation that injects spurious local magic, and is removed by error mitigation~\cite{Cai2023,Ahmad2024}; the depolarizing controlled-Z (CZ) channel acts non-locally, leaving the non-local magic structurally intact while posing the principal obstacle to local erasure. Recognizing this correspondence is what makes the no-free-parameter noise model possible.
This correspondence turns non-local magic into a benchmark that reaches where gate-fidelity metrics cannot. Randomized Benchmarking and its variants quantify Clifford gate fidelities~\cite{Magesan2011,Magesan2012,Chen2022}, and recent extensions incorporate non-Clifford gates~\cite{Cross2016,Helsen2019}, yet none reports on a device's ability to generate and manipulate magic for quantum advantage. Measuring local and non-local magic supplies exactly this missing information, offering an accessible, hardware-aware diagnostic for gate calibration and for benchmarking the computational resources behind quantum advantage.

\section*{Results}

\subsection*{Local and non-local stabilizer entropy}
Stabilizer entropy characterizes the extent to which a quantum state spreads over the Pauli operator basis, serving as a measure of its departure from stabilizer states~\cite{Leone_Oliviero_Hamma_2022}. The $\alpha-$SRE is defined as
\begin{equation}
\begin{split}
    M_{\alpha} (\psi) = \dfrac{1}{1 -\alpha} \log_{2}{ \left( \dfrac{1}{d} \sum\limits_{P \in \mathcal{P}_{N}} \vert {\rm Tr} (P\psi ) \vert^{2 \alpha} \right)},
    \label{magicformula}
\end{split}
\end{equation}
where $ \psi = \vert \psi \rangle \langle \psi \vert $ is the density matrix of the $N$-qubit pure state $\vert \psi \rangle$ of dimension $d = 2^{N}$. The set $\mathcal{P}_N$ represents the  $N-$qubit Pauli strings composed from the identity and Pauli operators $\{ \id,X, Y, Z \}$. For pure states, SRE with $\alpha \ge 2$ constitutes a proper magic monotone in the context of magic-state resource theory~\cite{Leone_Oliviero_Hamma_2022,Leone_Bittel_2024}. For mixed states, this quantity can be extended by subtracting the $2$-R\'enyi entropy, that is, $-\log_2(\pur (\psi))$ where $\pur(\psi)=\Tr(\psi^2)$ is the purity of $\psi$. 

The interplay between entanglement and non-stabilizerness is non-trivial: for example, maximal amount of SRE can only be achieved by highly entangled states~\cite{iannotti2025coventmag}. Moreover, one is interested in the non-stabilizer resources that cannot be extracted (or erased) by a local quantum protocol. This motivates 
non-local magic being defined as follows~\cite{Cao_Cheng_Hamma_Leone_Munizzi_Oliviero_2024}: given a bipartite system $\hi=\hi_A \otimes \hi_B$, then we define
\ba
M_2 ^{\rm NL}(\ket\psi) :=\min\limits_{U_A,U_B} M_2(U_A \otimes U_B \ket{\psi})
\label{eq:nlmagic}
\ea
that is, the minimal amount of non-stabilizerness contained in a bipartite state upon optimization in all local bases. 
Conversely, \textit{local} magic of a state is defined as the difference between the total magic and the non-local one $M_2 ^{\rm L}(\ket\psi)=M_2(\ket\psi)-M_2 ^{\rm NL}(\ket\psi)$ and corresponds to the amount of magic that \textit{can} be removed from a state by local gates. Notice that non-local magic is not simply magic in entangled states: there are entangled states that only host local magic, as in any state of the form $(U_A\otimes U_B) C\ket{0}$ where $C\ket{0}$ is an entangled stabilizer state. Remarkably, maximally entangled states only possess local magic. We denote the class of states achieving $M_2^{\rm NL}(\ket{\psi_\lambda})=M_2(\ket{\psi_\lambda})$ as \textit{non-local magic states} (NLM). To fix a bit the nomenclature, we will call \textit{local magic} (LM) states those that only possess magic locally, that is such that $M_2^{\rm NL}(\ket{\psi_\lambda})=0$. Finally, we will label simply by M those states such that $M_2(\ket{\psi_\lambda})>M_2^{\rm NL}(\ket{\psi_\lambda})>0$, that is, they possess both local and non-local SRE.

As we stated above, $M_2^{\rm NL}$ is also the amount of non-stabilizer resource that cannot be distilled by local unitary operations. To show it, we establish the following 
\begin{thm}[Local magic distillation]
    Given a state $\ket{\psi}\in \hi=\hi_A\ot \hi_B$, an ancillary system in $\ket{0}$, and a Clifford unitary $C$ such that
    \begin{equation}
        C(\ket{\psi}\ot \ket{0})=\ket{\psi'}\ot \ket{\phi}
    \end{equation}
    with $M_2(\ket{\phi})>M_2^{\rm L}(\psi)$, then $C$ does not allow a decomposition in local unitary operators on $\hi_A\ot \hi_B$, namely it cannot be of the form $C=C_A\otimes C_{BC}$ or $C=C_{B}\otimes C_{AC}$.
   \end{thm}
   \noindent As one can see, if $\ket{\psi}$ is a non-local magic state and $C$ is local, then $M_2(\ket{\phi})=0$, that is, we cannot distill any non-stabilizerness. 
The proof is  shown in Supporting Information.

Despite being the result of an optimization procedure, non-local magic is analytically computable for pure two-qubit states~\cite{Qian_Wang_2025} and is a function of the purity of the reduced density matrix (RDM) $\psi_A=\Tr_B\ketbra{\psi}$, namely $\pur(\psi_A)=\Tr(\psi_A^2)$, and one can show that 
\begin{equation}
    M_2^{\rm NL}(\ket\psi)=-\log_2 \left(4 \pur(\psi_A)^2-6 \pur(\psi_A)+3\right),
\end{equation}
see SI for details. 
The closed form above measures non-local magic from the anti-flatness of the entanglement spectrum, in the spirit of Ref.~\cite{Cao_Cheng_Hamma_Leone_Munizzi_Oliviero_2024}: the deviation of the spectrum from flatness, encoded in the RDM purity, is precisely what quantifies the non-local magic. Measuring the purity therefore yields a {\em state-agnostic} estimate of non-local magic. This route is in principle scalable, every measure of anti-flatness \cite{Cao_Cheng_Hamma_Leone_Munizzi_Oliviero_2024}, for instance capacity of entanglement allows for a lower bound to non-local SRE.  The second route is  state-dependent, finding the optimal local basis from Eq.~\ref{eq:nlmagic} and measure the residual SRE in the full state; this route too can be made state-agnostic by optimizing over $U_A,U_B$ via gradient descent. The two methods complement and cross-check each other.

In the following, we measure non-local magic both from the local optimal erasure experimental protocol via local unitaries (obtained numerically knowing state preparation) and from the state-agnostic subsystem purity measurements. Taken together, these two approaches provide a \emph{bona-fide} experimental demonstration of non-local magic and of the capabilities of addressing it.

\subsection*{Experimental Protocol}

\label{experimental_results}

\begin{figure*}[t!]
\centering
\includegraphics[width=\linewidth]{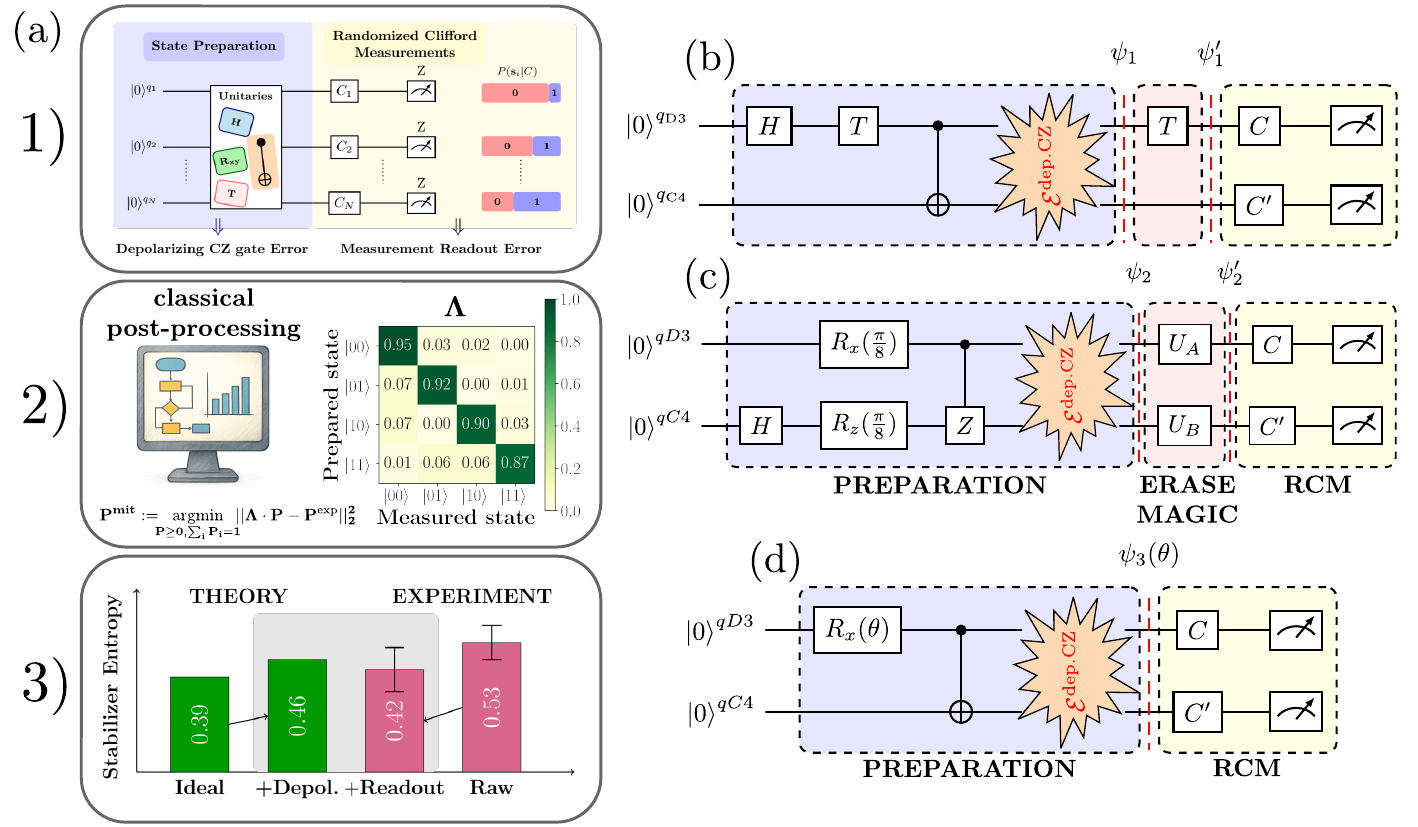}
\caption{Experimental approach and investigated circuits.
(\textbf{a}) Experimental approach for SRE measurement and error mitigation. 
The studies of LM, M, and NLM quantum states involve: (1) preparing target states and performing standard benchmarking protocols to extract CZ gate fidelity and readout calibration matrix $\Lambda$, (2) using $\Lambda$ in classical post-processing to obtain error-mitigated probabilities $P(\mathbf{s}_i|C)$, and (3) adjusting theoretical predictions for stabilizer entropy and purity given the observed CZ depolarizing error in case of the M state $\psi_2$.  
(\textbf{b}–\textbf{d}) Investigated circuits. The first two experiments comprise preparation and erasure protocols, while the third involves only non-local magic, confirmed by RDM-purity-inferred data (see Fig.~\ref{plot_nlm_exp}(b)). In all panels, the depolarizing CZ error $\mathcal{E}_p^{\mathrm{dep.\ CZ}}$ is accounted for at the theoretical-prediction level. \vspace{-0.5cm}}
\label{fig:NoiseModel_and_LMNLM}
\end{figure*}
As we have stated above, one can measure $M_2^{\rm NL}$ in essentially two ways: (a) state-dependent: one finds the optimal local erasure protocol $U_A\otimes U_B$ and then measures the full SRE. (b) state-agnostic: one measures spectral quantities of the RDM to directly evaluate $M_2^{\rm NL}$. 
We perform measurements on NLM, LM and M states. The NLM and M states host non-local magic   $M_2^{\rm NL}$ while LM states have $M_2^{\rm NL}=0$. The latter experiment is needed to show that one can effectively find the optimal local basis in which the state becomes a stabilizer state, that is, one can locally erase all the SRE.  
Figure~\ref{fig:NoiseModel_and_LMNLM}(a) summarizes the measurement scheme, common to the LM, M, and NLM experiments, which proceeds in three stages. First, the target state is prepared by the circuit of interest [Fig.~\ref{fig:NoiseModel_and_LMNLM}(b-d)]; in the local-erasure protocol the preparation is followed by the single-qubit unitaries that remove SRE locally, so that only the residual, non-local magic remains to be measured. Second, the state is characterized by randomized Clifford measurements (RCM): we sample random Clifford unitaries $C$, apply each to the prepared state, and read out in the computational basis, accumulating the outcome distribution $P(\mathbf{s}|C)$ over $N_{\rm shot}$ shots per Clifford. Both the stabilizer R\'enyi entropy $M_2$ and the subsystem purity are estimated from the statistics of $P(\mathbf{s}|C)$ across the sampled Cliffords (see Methods), so that a single dataset feeds both the local-erasure and the state-agnostic purity routes. Third, the data are classically post-processed: readout errors are mitigated with the calibration matrix $\Lambda$ obtained from device benchmarking, giving the corrected probabilities $P(\mathbf{s}_i|C)$, while on the theory side the dominant intrinsic noise, a depolarizing channel $\mathcal{E}_p^{\rm dep.\,CZ}$ acting on the native CZ gate, is folded into the predictions for $M_2$ and purity with a strength fixed by the independently measured interleaved randomized benchmarking (IRB) CZ fidelity. Both corrections are parameter-free: $\Lambda$ and the CZ depolarizing strength come from separate benchmarking experiments, so the comparison between theory and experiment involves no fitting.

Let us now describe the three specific quantum circuits preparing LM, M and NLM states, shown in Fig.~\ref{fig:NoiseModel_and_LMNLM} (b-d). 

\textit{\textbf{Local Magic (LM) state.---}} We first consider a circuit that only hosts local magic injected by a $T$ gate, despite the state being maximally entangled (see Fig.~\ref{fig:NoiseModel_and_LMNLM}(b)). The preparation circuit output reads $\psi_1 =\mathcal{E}_p ^{\rm dep. CZ} [2^{-1/2} (\ket{00} + e^{i \frac{\pi}{4}} \ket{11})]$. Evaluation of stabilizer entropy for this state yields $M_{2} (\psi_1) \sim 0.48$. One removes or erases local magic in the state by applying another $T$ gate on the same qubit, yielding $M_{2} (\psi'_1 ) \sim 0.1$. This constitutes a minimal example where we are able to (almost) completely remove magic with a local gate. 
We observe consistency between theoretical expectations and experimental results for both purity and stabilizer entropy within the statistical error associated with the RCM, together with the results obtained by the measurement of RDM purity, as seen in Table~\ref{tab:experiment1}.

\textit{\textbf{Local+Non-Local  Magic (M) state.---}} We measure the response of the quantum circuit in  Fig.~\ref{fig:NoiseModel_and_LMNLM}(c) for which the output state has both local and non-local magic. 
The output state of the preparation circuit reads
 $ \psi_2=\mathcal{E}_p ^{\rm dep. CZ}[ 2^{-3/2}(c_+\big(\ket{00}+e^{i\frac{\pi}{8}}\ket{10}\big)-i c_-\big(\ket{01}+e^{i\frac{\pi}{8}}\ket{11})\big)]$, 
with $c_{\pm}=(2\pm(2+\sqrt{2})^{\frac12})^{\frac12}$, and stabilizer entropy is  $M_2(\psi_2)\sim 0.46 $. Since this state has both local and non-local magic, one expects to be able to partially erase magic using single-qubit gates. Numerical optimizations yield the following local magic-erasing unitaries: $ U_A=R_z(\alpha)R_y(\beta)R_z(\gamma)\;$ and $U_B=R_z(\delta)R_y(\eta)R_z(\phi)\,,$ with the rotation angles $\alpha = \beta = \gamma = \delta = \eta = 0$ and $\phi = 67.61^\circ$. Experimental data strongly validates theoretical predictions, as summarized in Tab.~\ref{tab:experiment1}.

\textit{\textbf{Non-Local Magic (NLM) states.---}} We consider a two-qubit state of the form $\psi_3 (p,\!\theta)\!=\mathcal{E}_p ^{\rm dep. CZ}\!\left[ \cos( \theta/2) \!\ket{00} \!- \!i \!\sin( \theta /2)  \!\ket{11}\!\right]$, realized by the quantum circuit in Fig.~\ref{fig:NoiseModel_and_LMNLM}(d). This state only hosts non-local magic, and including the depolarizing CZ noise channel we obtain the following closed-form expression for the stabilizer entropy $ M_{2} ( \psi_3 (p,\theta) ) = -\log_2 \big[4 \big((p-1)^4 \cos (4 \theta ) + 5 (p-2) p ((p-2) p + 2) + 7 \big)\big]  + \log_2 (3 (p-2) p+4) + 3$,
\begin{figure}[t!]
\centering
\includegraphics[width=0.9\columnwidth]{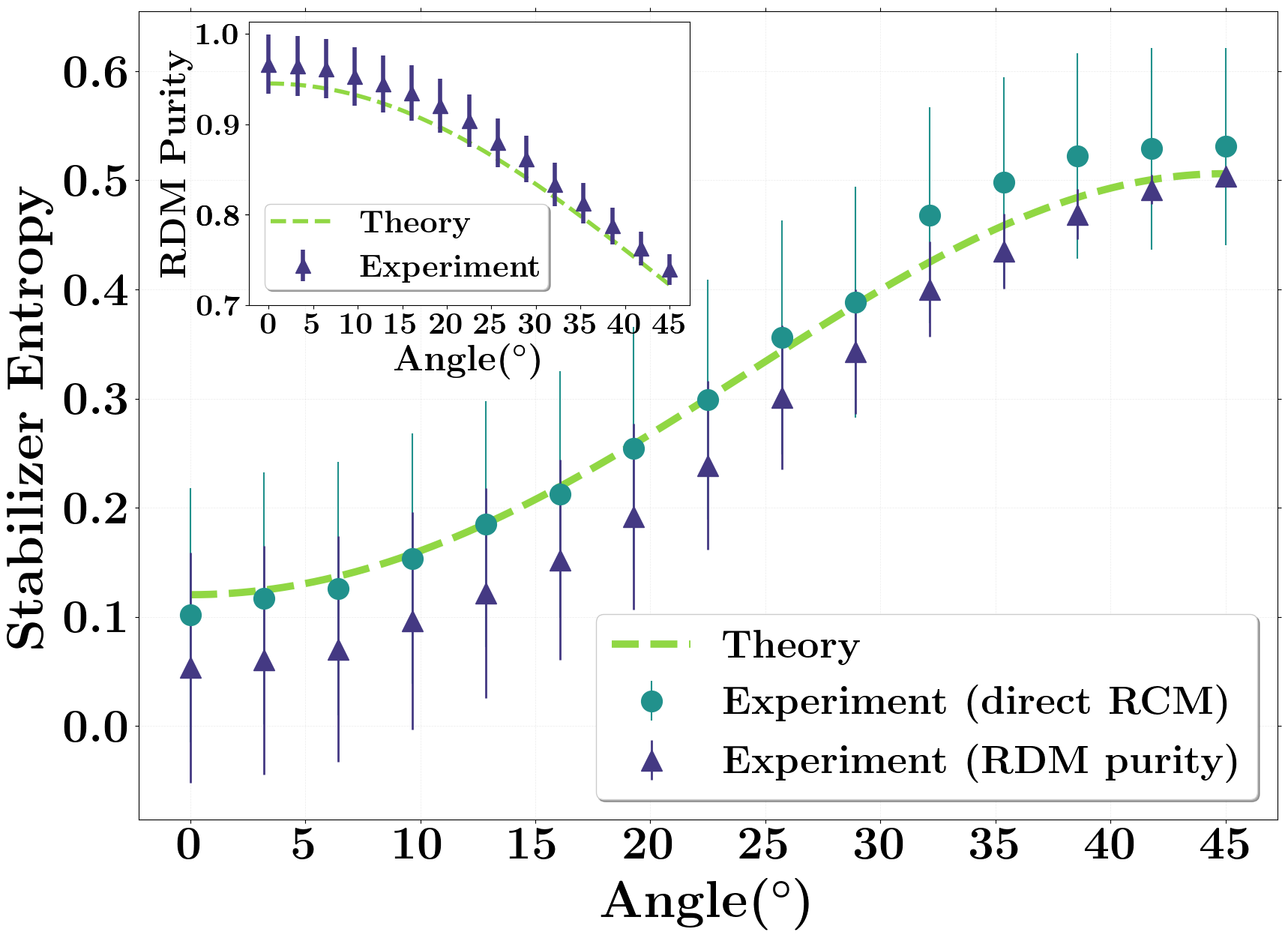}
\caption{Experimental readout-error-mitigated data  as well as the values of non-local magic obtained by RDM purity compared to the theoretical line of the stabilizer entropy and RDM purity (inset) with CZ depolarizing errors taken into account.\vspace{-0.5 cm} }
\label{plot_nlm_exp}
\end{figure}
A systematic investigation of the stabilizer entropy as a function of $\theta \in[0,\pi/4]$ is shown in Fig.~\ref{plot_nlm_exp}. 

\begin{table*}[t!]
\caption{\label{tab:experiment1} Comparison of purity and magic. Obtained values for states $\psi_i$ (before) and $\psi'_i$ (after) local magic erasure for $i= \{1,2\}$, as well as the value of non-local magic obtained from subsystem purity by tracing out qC4 qubit. The theoretical predictions for purity and magic are calculated by considering the CZ depolarizing error and compared with the experimental results after readout error mitigation.
}
\begin{tabular}{@{\extracolsep{\fill}}cccccccc}
\hline\hline
\textbf{Class} & \textbf{Circuit} & \textbf{State} & \textbf{Th. Purity} & \textbf{Exp. Purity} & \textbf{Th. Magic} & \textbf{Exp. Magic} & \textbf{NL-Magic (from RDM purity)} \\
\multirow{2}{*}{\centering Local Magic (LM) }  & \multirow{2}{*}{Fig.~\ref{fig:NoiseModel_and_LMNLM}(b)} & $\psi_1$ & 0.94 & $\textbf{0.90}\pm 0.04$ & 0.48 & $\textbf{0.38}\pm 0.09$ & 
\multirow{2}{*}{\centering $\textbf{0.0888}\pm 0.0004$} \\
 & & $\psi'_1$ & 0.94 & $\textbf{0.93}\pm 0.05$ & 0.08 & $\textbf{0.1}\pm0.1$ & \\
 \hline
\multirow{2}{*}{\centering Local+Non-Local (M) } & \multirow{2}{*}{Fig.~\ref{fig:NoiseModel_and_LMNLM}(c)} & $\psi_2$ & 0.94 & \textbf{0.94}$\pm 0.04$ & 0.46 & $\textbf{0.42}\pm 0.09$ & 
\multirow{2}{*}{\centering $\textbf{0.19}\pm 0.07$} \\
 & &$\psi'_2$ & 0.94 & $\textbf{1.00}\pm 0.05$ & 0.27 & $\textbf{0.3}\pm0.1$ & \\
\hline\hline
\end{tabular}
\end{table*}

\begin{figure}[ht!]
\begin{center}
        \includegraphics[width=\linewidth]{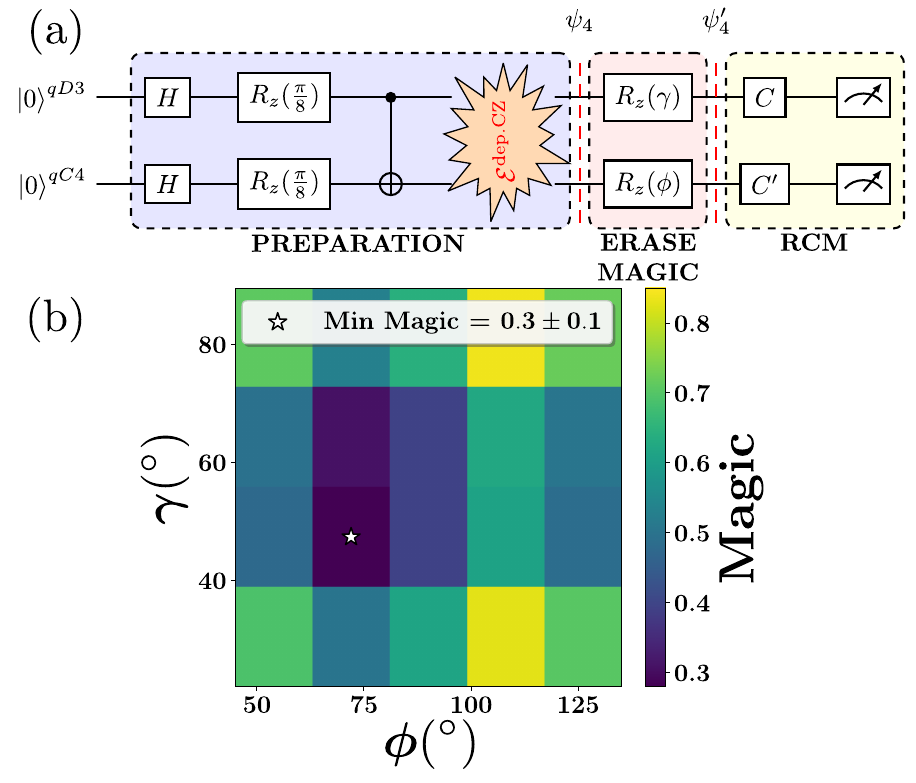}
        \caption{Quantum circuit for Local+Non-local magic quantum state with CNOT gate. Panel (a): circuit representation of the sweeping protocol for empirically estimating the optimal magic erasure angles. The minimum amount of theoretically obtainable magic using $R_z$ rotations is $0.29$ with $(\gamma,\phi)=(67.5^\circ,90^\circ)$. Panel (b): results, with the star marking the values of the angles for which the state has minimal magic. The sweeping angles method reaches the same minimum value as theoretically predicted, although for different angles than anticipated. \vspace{- 1cm} 
        }
    \label{cplx_NLM}
    \end{center}
\end{figure}

\section*{Materials and Methods}

\subsection*{Device characterization and benchmarking}
The initialization experiment involves the preparation of an $N$-qubit computational state $\ket{i}$. We then simultaneously acquire a readout voltage across each qubit readout resonator in the $I,Q$ plane, and identify the state discrimination threshold as discussed in Ref.~\cite{Stasino2025} by using the conditional probability method therein introduced. This also allows us to obtain the readout probability matrix, i.\,e. the probability to measure the state $\ket{j}$ over the computational basis states as the ratio between the number of counts in the different basis states $n^{\rm init}_{ij}$ and the total number of shots $N_{ \rm shot}$, $p^{\rm init}_{ij}=n^{ \rm init}_{ij}/N_{\rm shot}$. The readout probability matrix is finally used to calculate the readout fidelity per quantum circuit. In this work, we have used $N_{\rm shot}=5000$ for determining the outcome probability distributions $P(\textbf{s}_l|C)$ for each given Clifford pair $C$.

After a thorough characterization and benchmarking of the qubits, including measurements of readout and gate fidelities, coherence times, and crosstalk effects, we performed measurements of the SRE for the quantum states summarized in SI Appendix Fig. S2 (c-f). The corresponding experimental results are presented in Table~S1, which reports the measured fidelities, coherence, and crosstalk parameters. These data provide direct insight into how the relevant sources of error affect the estimation of the SRE in our device.  

We begin by considering single-qubit states prepared on both the D3 and C4 qubits. Specifically, we examine the computational basis state \( \ket{0} \) and the superposition states \( \ket{+} \) and \( \ket{-} \), obtained by initializing the qubit in either \( \ket{0} \) or \( \ket{1} \) and subsequently applying a Hadamard gate, as illustrated in SI Appendix Fig. S2 (c) and (d).  
In these experiments, we expect vanishing SRE. By contrast, when a \( T \) gate is introduced, the experimentally measured SRE increases and approaches the theoretical value of \( \log_{2}(4/3) \approx 0.41 \).  
Next, we measure the SRE for two-qubit quantum states according to the circuit depicted in SI Appendix Fig. S2 (e).  At this stage, we choose not to implement any entangling gate between the two qubits. Since the SRE is additive, experimental results demonstrate that RCM can obtain experimental magic values consistent with the sum of the values measured in the isolated case, both with and without the injection of non-Clifford gates. Additivity is consistent with the fairly low impact of microwave and ZZ crosstalk~\cite{Zhao2022,Ketterer2023}.

Finally, we measure the SRE for a Bell state, thus including more complexity and hardware noise sources in the experiment (see SI Appendix Fig. S2 (f)). From SI Appendix Table S1, one notices: i) readout fidelities are lower in the case of two-qubit states; ii) interleaved RB fidelities for the native two-qubit gate CZ is considerably lower than single-qubit gate RB fidelities, either measured in the isolated (a) and coupled scheme (b), thus implicitly including any microwave or ZZ-crosstalk. This explains the presence of non-zero experimental SRE in those cases where one does not expect any.

We employ a classical post-processing error mitigation scheme to counteract the effect of readout errors. This scheme directly uses information from different benchmarking tools to single out sources of crosstalk, thanks to direct experimental control of the chip. In particular, our scheme relies on the knowledge of the readout calibration matrix (see SI Appendix Sec. 3A). This makes the error mitigation procedure ``parameter free'', as it does not require any fitting parameter across a vast set of experiments, as done in other works, such as Ref.~~\cite{Oliviero_Leone_Hamma_Lloyd_2022}.

In our hardware, CNOT is decomposed in terms of a combination of single-qubit Hadamard gates and the CZ gate. Motivated by the relatively larger CZ errors on the measured purity and stabilizer entropy, we account for its influence on the theoretical prediction side by modeling the effects as a depolarizing channel (SI Appendix Sec. 3A), with a depolarization strength directly based on the IRB CZ fidelity. Our approach, allowing for a consistent comparison between theoretical expectations and the experimental outcomes, is summarized in Fig.~\ref{fig:NoiseModel_and_LMNLM} (a). 

\subsection*{The device}
\label{sQPU}

The superconducting Quantum Processing Unit (sQPU) used in this work is a Contralto-D from Quantware (SI Appendix Fig. S2 (a)), composed of $21$ flux-tunable transmons and $4$ fixed-frequency transmon qubits (in yellow). The fixed-frequency qubits are specifically designed for coherence times benchmarking, i.\,e. they lack drive lines to minimize radiative interaction with the environment, are isolated and not affected by flux noise. The sQPU is thermally and mechanically anchored at the coldest stage of a BlueforsXLD1000SL (the mixing chamber, MXC), and is enclosed by two magnetic shields. At room temperature, the system uses Qblox electronics, which is comprised of two main parts: Clusters (equipped with RF modules to implement drive, readout, and two-qubit gates) and a low-noise SPI rack for setting static DC flux.  
The full description of the setup is reported in the Supporting Information (SI).

The $21$-qubits matrix has rectangular 2D connectivity, where high-frequency bus resonators (in grey) are used to couple neighboring qubits~\cite{DiCarlo2009}. Qubits are designed to fall in three frequency bandwidths: high-frequency qubits (in red) have frequencies of the order of $6\;GHz$, intermediate frequency qubits (in blue) have frequencies of the order of $5\;GHz$, low frequency qubits (in green) lie in a frequency range of the order of $4\;GHz$. Each qubit is equipped with a dedicated superconducting readout resonator, equally distributed over $4$ readout feedlines in a notch-type geometry (highlighted with different colors for clarity), hence guaranteeing multiplexed readout. Dedicated drive and flux lines are used to implement X, Y, Z single-qubit and two-qubit gates, respectively~\cite{Krantz2019}. In fact, flux-lines are used to implement the native two-qubit gate of the processor, the Controlled-Z (CZ) gate,  using magnetic Sudden-net-Zero (SNZ) flux pulsed signals~\cite{Negirnac2021}, combined with low $1/f$-noise static flux fields through bias-tees at room-temperature. Flux lines also allow for parking qubits at specific working points by using a static magnetic flux. 

In this work, we focus on qubits D3 and C4 (intermediate and low frequency qubits), operated at their flux sweet spots (SI Appendix Fig. S2 (b)). Additionally, the neighboring qubit A6 has been placed at its sweet spot (at least $1GHz$ detuned from the interested qubits), whereas the remaining qubits of the matrix are operated at their anti-sweet spot (in black). This allows us to operate in a quasi-ideal $2$-qubit register and avoid frequency crowding. Notably, flux-tunable transmons in the processor are designed to be symmetric~\cite{Koch2007}.

\subsection*{Randomized Clifford Measurements}
\label{RandomizedToolbox}

To overcome the exponential cost of direct measurement of SRE with qubit number, we use a more scalable protocol, namely Randomized Clifford Measurements (RCM)~\cite{Oliviero_Leone_Hamma_Lloyd_2022}.
Anticipating purity-impacting noise in our device, we use the mixed-state extension of SRE.
Thus, our experimental approach consists of two main steps: i) preparation of an \( N \)-qubit state \( \psi \); ii) implementation of the RCM protocol to evaluate purity \( \mathcal{P} \) and stabilizer purity \( \mathcal{W} \) of the prepared state \( \psi \). These two independent quantities enable estimation of  SRE according to  
\begin{align}
\label{SRE}
M_{2} (\ket{\psi}) = -\log_{2} \mathcal{W} + \log_{2} \mathcal{P} - \log_{2} d,
\end{align}
where $\mathcal{P} = d \sum_{\vec{\mathbf{s}}} (-2)^{-\| \vec{\mathbf{s}} \|_{2}} \, \mathbb{E}_{C} \big[ P(\mathbf{s}_1 \vert C) P(\mathbf{s}_2 \vert C) \big]$,  and $\mathcal{W} = - \sum_{\vec{\mathbf{s}}} (-2)^{-\| \vec{\mathbf{s}} \|_{4}} \, \mathbb{E}_{C} \big[ P(\mathbf{s}_1 \vert C) P(\mathbf{s}_2 \vert C) P(\mathbf{s}_3 \vert C) P(\mathbf{s}_4 \vert C) \big]$. Here, \( \mathbf{s}_i \) denote computational basis strings, and \( \| \cdot \| \) refers to their corresponding Hamming weight. The expectation value \( \mathbb{E}_{C} \) indicates averaging over different random choices of single-qubit Clifford gates \( C \). To ensure statistical reliability, a sufficiently large number of samples must be considered, as discussed in Ref.~\cite{Oliviero_Leone_Hamma_Lloyd_2022}.

The RCM protocol exploits the use of a randomized set of Clifford unitaries to estimate, by sample average, the purity and the stabilizer purity. The cardinality of the possible $N$-qubit Clifford unitaries (modulo global phase $U(1)$)~\cite{Grier_Schaeffer_2022} that are randomly picked from the Clifford group reads as
\begin{align}
    \vert \overline{C_N} \vert = \vert C_N/ U(1) \vert = 2^{N^2 + 2N} \prod\limits_{k = 1}^N \left( 4^k - 1 \right).
\end{align}
Since the number of possible Clifford combinations for a single-qubit is not particularly demanding, being $\vert \overline{C_1} \vert = 24$, in this work, we have used all the possible unitaries to experimentally assess the fundamental quantities cited above for single-qubit states. For two-qubits, instead, the number of possible products of single-qubit Clifford unitaries dramatically increases ($\vert \overline{C_1\otimes C_1} \vert = 576$) and in general the cardinality grows as $24^N$. To reduce the experimental effort in calculating two-qubit states and purities, we have randomly picked $N_{ \rm rand}=400$ Clifford gates from the $C_2$ gate set, where $N_{ \rm rand}$ is the number of randomizations. 

For each choice of single-qubit Clifford $C$, we measure the readout probability vector $P(\textbf{s}|C)$. The vector components of $P(\textbf{s}|C)$ are obtained as the number of events $n_{ \textbf{s}}$ for which the measured output corresponds to the computational basis states represented by the bitstring $\textbf{s}$  normalized to the total number of shots measurements $N_{ \rm shot}$, $p_{\textbf{s}}=n_{\textbf{s}}/N_{ \rm shot}$. The state assignment process uses the discrimination threshold obtained by an initialization experiment, unique for each quantum state preparation (as detailed in SI Appendix Sec. 1). 

\section*{Discussion and conclusions}

Genuine quantum advantage requires the coexistence of non-stabilizerness and entanglement, an interplay captured by non-local magic~\cite{Cao_2024_area,Odavic_Haug_Torre_Hamma_Franchini_Giampaolo_2023}. Here we have provided its first experimental measurement on a QPU.

There is at present no single, agreed-upon way to benchmark QPU performance across the many circuit designs and error-mitigation schemes in use, so connecting the manipulation of magic on NISQ hardware to a device's physical characteristics and noise sources is a pressing challenge. The processor studied here, the core of the first Italian Superconducting Quantum Computing Center (\emph{Partenope}), is well suited to it: direct experimental control allows an error-mitigation approach tailored to the measured performance of the device. This was key both to demonstrating local magic erasure and to observing non-local magic through subsystem purity.

Our hardware-informed noise model identifies measurement readout and CZ depolarizing noise as the dominant error sources. Readout error is chiefly responsible for injecting spurious local magic, whereas the depolarizing channel leaves the non-local structure of magic intact. These are not the only noise sources in our device, and more sophisticated mitigation may be required for State Preparation and Measurement (SPAM) errors as circuit depth grows. Importantly, erasing local magic does not require full knowledge of the noise channels affecting the chip, as we show next.

We demonstrate a proof-of-concept protocol that identifies, empirically and in situ, the unitaries that erase local magic (Fig.~\ref{cplx_NLM}). Theory restricts the search to a small family within which the optimal erasing unitaries must lie; sweeping the rotation angles over the relevant interval then isolates gates that remove the local magic [Fig.~\ref{cplx_NLM}(b)]. The minimum reached closely matches the theoretical value for the state, even though the optimal angles differ from the predicted ones. This discrepancy points to noise sources beyond our model, and underscores that the local magic-erasing unitaries can be found heuristically, without a complete noise characterization.

In short, any study of local and non-local magic must reckon with the noisy nature of the hardware. Our results open a route to benchmark and calibrate quantum devices beyond traditional gate-fidelity protocols, one we expect to carry over to other architectures and platforms. They also bring within experimental reach a range of systems in which non-local magic is central, from the violation of Bell inequalities in NLM states~\cite{cusumano2025nonstabilizernessviolationschshinequalities} to laboratory toy models of black holes.

These capabilities also underpin a concrete metrological advantage, and clarify where its cost resides. Because local unitaries leave the entanglement spectrum, and therefore the subsystem purity, invariant, the local magic our protocol erases is spectrally inert: a state with no non-local magic has a flat entanglement spectrum, so its purity is fixed by the entanglement entropy alone, and it is the non-local magic that sets the spectral structure a purity estimator must resolve. The cost of estimating purity is thus governed by the non-local content of a state rather than by its total non-stabilizerness, and erasing local magic in situ, as we do here, is the elementary instance of the cleansing step that strips away the free, local part. This is the mechanism behind the efficient purity-estimation algorithm of Ref.~\cite{Leone_Oliviero_Esposito_Hamma_2024}, which attains an exponential advantage over state-of-the-art protocols; a complete characterization of purity-estimation cost in terms of non-local magic will be presented in forthcoming work. The same algorithmic kernel decodes Hawking radiation from a black hole without prior knowledge of the initial state~\cite{Leone_Oliviero_Lloyd_Hamma_2024,Oliviero_Leone_Lloyd_Hamma_2024_PRL,Leone_Oliviero_Piemontese_True_Hamma_2022}. Our experiment realizes both of its ingredients at the two-qubit level, reading non-local magic from purity and removing the local part in situ; with sufficient control over non-stabilizerness, we propose that such algorithms can be run on real quantum hardware.

\acknowledgements{The authors thank Lorenzo Campos Venuti for useful discussion, Hany Ali, Alessandro Bruno, Johannes Hermann, and Christian Junger for support on the gate calibrations. The work is supported by the PNRR MUR project PE0000023-NQSTI and the PNRR MUR project CN 00000013-ICSC. }

\subsection*{DATA AVAILABILITY}
The raw data, the data processing scripts and post-processed data are publicly available via Zenodo platform at \href{https://doi.org/10.5281/zenodo.17394953}{https://doi.org/10.5281/zenodo.17394953}. 

\subsection*{CODE AVAILABILITY}
All data processing codes are publicly available via Zenodo platform at \href{https://doi.org/10.5281/zenodo.17394953}{https://doi.org/10.5281/zenodo.17394953}.

\appendix

\onecolumngrid

\section{Initialization experiment}
\label{init}

The initialization experiment involves the preparation of an $N$-qubit computational state $\ket{i}$. We then acquire simultaneously a readout voltage across each qubit readout resonator in the $I,Q$ plane, and identify the state discrimination threshold as discussed in Ref.~\cite{Stasino2025} by using the conditional probability method therein introduced. This also allows us to obtain the readout probability matrix, i.\,e. the probability to measure the state $\ket{j}$ over the computational basis states as the ratio between the number of counts in the different basis states $n^{\rm init}_{ij}$ and the total number of shots $N_{ \rm shot}$, $p^{\rm init}_{ij}=n^{ \rm init}_{ij}/N_{\rm shot}$. The readout probability matrix is finally used to calculate the readout fidelity per quantum circuit. In this work, we have used $N_{\rm shot}=5000$ for determining the outcome probability distributions $P(\textbf{s}_l|C)$ for each given Clifford pair $C$.

\subsection*{Randomized Clifford Measurements experimental parameters and errors}

Since $\mathbb{E}_{C}$ represents the \emph{sample mean} of $P(\textbf{s}_l \vert C)$ over the number of Clifford gates $N_C$, 
\begin{align}
	\bar{x} = \exv_C \left[ x \right] \equiv  \frac{1}{N_C} \sum_{i=1}^{N_C} x_i\,,
\end{align}
where $x$ either represents purity $\mathcal{P}$ and stabilizer purity $\mathcal{W}$,  the  \emph{statistical sample variance} is calculated in terms of the square deviations from the mean $ d_i^2 = (x_i - \bar{x})^2$ as
\begin{equation}
	s^2 = {\rm Var} \left( x \right)= \exv_C \left[ \left( x - \bar{x} \right)^2 \right] \equiv \frac{1}{N_C - 1} \sum_{i=1}^{N_C} (x_i - \bar{x})^2.
\end{equation}
It follows that the \textit{statistical sample standard deviation} is the square root $s = {\rm Std} (x) = \sqrt{s^2}$, which represents, as usual, the spread of the experimental measurements. However, when taking a limited sample from a probability distribution, an inherent error is induced by the finite sampling. Therefore, we introduce the \textit{sampling error}, defined in terms of the statistical sample standard deviation $s$ as
\begin{align}
	\Delta x  = \frac{s}{\sqrt{N_C}}\,.
\end{align} 
Finally, given the non-linear (logarithm) function behavior of our random variables $\mathcal{P}$ (purity) and $\mathcal{W}$ (stabilizer purity), we are required to perform error propagation to get the average and statistical sampling error of Stabilizer Rényi entropy and 2-Rényi entropy. Using standard methods of error propagation, the estimator of the variance of a function $f$ of two random variables $x$ and $y$ is given by
\begin{equation}
	\begin{split}
		&\var [f(x,y)] \approx \left|\bigg(\frac{\partial{f}}{\partial{x}}\bigg)_{\bar{x},\bar{y}}\right|^2\var (x)\\ &+\left|\left(\frac{\partial{f}}{\partial{y}}\right)_{\bar{x},\bar{y}}\right|^2\var (y)+2\left(\frac{\partial{f}}{\partial{x}}\frac{\partial{f}}{\partial{y}}\right)_{\bar{x},\bar{y}}\!\!\!\!\!{\rm Cov}(x,y)\,,
	\end{split}
\end{equation}
which for the SRE $M_2$ yields
\begin{equation}
	\begin{split}
		\var(M_2)&\approx \frac{\var(\mathcal{W})}{(\bar{\mathcal{W}}\ln 2)^2}+\frac{\var(\mathcal{P})}{(\bar{\mathcal{P}}\ln 2)^2} +2\, \frac{{\rm Cov (\mathcal{W},\mathcal{P})}}{\bar{\mathcal{W}}\bar{\mathcal{P}}(\ln 2) ^2}\,.
	\end{split}
\end{equation}
However, in Ref.~\cite{iannotti2025coventmag}, it is shown that the covariance between $\mathcal{P}$ and $\mathcal{W}$ is exactly zero, so we can neglect that term in the expression, finally yielding
\begin{equation}
	\Delta M_2=\frac{1}{\ln 2}\sqrt{\frac{\var(\mathcal{W})}{N_C \bar{\mathcal{W}}^2}+\frac{\var(\mathcal{P})}{N_C \bar{\mathcal{P}}^2}}\,.
\end{equation}

\section{Non-local magic as a function of RDM purity}
\label{rdm}
In this section, we derive the expression of non-local magic as a function of RDM purity, valid in the case of two-qubit states.
Any pure two qubit state $\ket{\psi}$ can be written in the following way: $ \ket\psi=\sqrt{\lambda} \ket{\psi_1 ^A}\ket{\psi_1 ^B}+\sqrt{1-\lambda}\ket{\psi_2 ^A}\ket{\psi_2 ^B},$  where $\{\lambda,1-\lambda\}$ are called Schmidt coefficients and $\{\ket{\psi_1 ^X},\ket{\psi_2 ^X}\}$ is the Schmidt basis of $\hi_X$, with $X\in \{A,B\}$.
Given this expression, the non-local magic of $\ket\psi$ is equal to
\begin{equation}\label{nlschmidt}
	M_2^{\rm NL}(\ket\psi)=-\log_2 \left(4 (\lambda -1) \lambda  (1-2 \lambda
	)^2+1\right)\!,
\end{equation}
with states achieving $ M_2^{\rm NL}(\ket{\psi_\lambda})=M(\ket{\psi_\lambda})$ being of the form $\ket{\psi_\lambda}=\sqrt{\lambda}\ket{00}+\sqrt{1-\lambda}\ket{11}$, 
and $\ket{00},\ket{11}$ being states of the computational basis, i.e. the eigenstates of the $Z$-Pauli matrix. Note that Eq. \eqref{nlschmidt} reduces to the one derived in ~\cite{Qian_Wang_2025} when one parametrizes the Schmidt coefficients as $\{\cos(\theta/2),\sin(\theta/2)\}$, i.e.
\begin{equation}\label{mnltheta}
	M_2^{\rm NL}(\theta)=\log_2{ \left( 8 (7 + \cos{(4 \theta )}))^{-1} \right)}.
\end{equation}
Since non-local magic depends exclusively on the Schmidt spectrum, in the unique case of two-qubit states one is able to measure the Schmidt coefficients by measuring a single quantity, e.g. subsystem purity. Starting from a pure state $\ket\psi$ in the Schmidt form, the purity of the RDM $\psi_A:=\Tr_B{\ketbra{\psi}}$ is given by
$\pur(\psi_A)=\lambda^2+(1-\lambda)^2\,$. Inverting this relationship and substituting it in Eq. \eqref{nlschmidt}, we get an explicit function of non-local magic on RDM purity:
\begin{equation}\label{nlrdmpur}
	M_2^{\rm NL}(\ket\psi)=-\log_2 \left(4 \pur(\psi_A)^2-6 \pur(\psi_A)+3\right)\,.
\end{equation}
This method of deriving $M_2^{\rm NL}$ remains valid also in the presence of noise, although the expression will be different than the one shown in \eqref{nlrdmpur}. More specifically, the relationship between $\pur(\psi_A)$ and $\lambda$ will change according to the action of the specific noise channel, and will depend on the noise parameters $\va p_{\rm noise}$ as well. A detailed discussion on noise analysis is reported in Sec.~\ref{sec:noise}. Nevertheless, again inverting such relation and obtaining $\lambda(\pur(\psi_A),\va p_{\rm noise})$ and plugging it in Eq. \eqref{nlschmidt} one obtains an expression of $M_2^{\rm NL}$ as a function of $\pur(\psi_A)$ (and $\va p_{\rm noise}$, that need to be determined experimentally).

\section{RDM purity from global measurement data}
In case of the direct method for measuring non-local magic, we measure the subsystem purity $\pur(\psi_A)$ with $\psi_A:=\Tr_B (\ketbra{\psi})$ to extract the value of the Schmidt coefficient. Analogously to the global system purity, partial purity can be measured via Randomized Clifford Measurements using the following formula~\cite{Brydges_Elben_Jurcevic_Vermersch_Maier_Lanyon_Zoller_Blatt_Roos_2019}: 
\begin{equation}
	\pur(\psi_A)=\exv_C \,2^{N_A} \!\!\!\sum_{s_A, s_A^{\prime}}\!\!(-2)^{-\vert \vert\vec{\textbf{s}}_A\vert\vert_2} \!P\left(\textbf{s}_A|C\right) \!P\left(\textbf{s}_A^{\prime}|C\right),
\end{equation}
where the authors  used generic randomized unitaries to compute this quantity. However, the same result holds if one restricts the unitary gate set to be taken from the Clifford group, as it is a unitary 3-design ~\cite{Webb_2016}. The prefactors in terms of Hamming weight of strings $\vert \vert \vec{\textbf{s}} \vert \vert$ are defined as $\vert \vert \vec{\textbf{s}} \vert \vert_2 = \sum_{i = 1}^{n} s_{1}^{i} \oplus s_{2}^{i}$ where $\vert \vert \vec{\textbf{s}} \vert $ where the $\oplus$ symbol represents the modulo 2 sum or XOR between the bit strings. Note that in the RCM protocol the expression for evaluating the stabilizer purity involves the prefactor $\vert \vert \vec{\textbf{s}} \vert \vert_4$, involving XOR operations between 4 bit strings. 

One recovers the subsystem probability outcomes as the marginal probability distributions of the global measurement outcomes, thereby using the same data taken to measure the purity non-stabilizerness of the global state, as seen in the following calculation:
\begin{equation}
	\begin{split} P_\psi(\textbf{s})&=P(\textbf{s}_A,\textbf{s}_B)=\Tr[\psi\ketbra{\textbf{s}_A \textbf{s}_B}]\\ &=\Tr_A\{\Tr_B[(\id_A\otimes \ketbra{\textbf{s}_B})\psi]\ketbra{\textbf{s}_A}\}\\
		&=\Tr_A[\psi_A \ketbra{\textbf{s}_A}]\Tr_B[\psi_B \ketbra{\textbf{s}_B}]\\&=P_{\psi_A}(\textbf{s}_A)P_{\psi_B}(\textbf{s}_B)\,.
	\end{split}
\end{equation}
Therefore, summing over all possible outcomes of one of the two qubits we reconstruct the partial probability distribution on the other qubit as
\begin{equation}
	P_{\psi_A}(\textbf{s}_A)=\sum_{\textbf{s}_B}P(\textbf{s}_A,\textbf{s}_B)\,.
\end{equation}

\subsection{Noise analysis}\label{sec:noise}
In this section, we explain the main sources of noise by which the chip is affected, and the techniques we used to mitigate and obtain an appropriate theoretical error model for the experimental data. We created a model without making use of free parameters, and only using independent benchmarking data of the quantum processor. 
%As mentioned in the main text, 
We focused on two main sources, namely readout error in computational basis measurements, and depolarizing error on the $\cnot$ (or rather, the $\rm CZ$) gate, each of which we will tackle in a separate subsection.

%\subsection{Measurement readout}
%\subsection{Depolarizing Control+Z noise}

\textit{\textbf{Readout noise}} Readout error means the expected outcomes of measurements do not necessarily correspond to the actual preparation. For example, if we prepare a two qubit state in $\ket{00}$ and we find it in $\ket{10}$ upon measurement, it means a readout error has occurred.
The readout errors we encounter are of classical type, and come in different varieties: i) \emph{uncorrelated} (no cross-talk); ii) \emph{correlated}, which signals crosstalk, i.e., unwanted correlation between qubits existing on the quantum chip. 

Formally, a quantum measurement is modeled through the use of Positive-Operator Valued Measurements (POVM), which are positive-semi-definite operators $\Pi_x$, one for each possible measurement outcome, that satisfy $\sum_x \Pi_x=\Id$. The probability of getting an outcome $x$ on a certain state preparation $\rho$ is then given by the Born rule, namely $p_x={\rm Tr}[\Pi_x \rho]$. In the case of two qubits and a measurement in the computational basis, we have four POVMs, namely $\mathbf{M}^{\rm ideal}=(\Pi_{00},\Pi_{01},\Pi_{10},\Pi_{11})^T$, with $\Pi_{ij}= \ketbra{ij} $. 

However, in the presence of readout noise, the measurement actually performed is a noisy POVMs, denoted $\mathbf{M}^{\rm noisy}$, related to the ideal one through a left-stochastic matrix $\Lambda$, in the following way:
\begin{equation}
	\mathbf{M}^{\rm noisy}=\Lambda\cdot \mathbf{M}^{\rm ideal}\,,
\end{equation}
where $\Lambda$ is usually referred to as \textit{calibration} or \textit{noise matrix}. Practically, the presence of readout noise induces a transformation on the would-be ideal outcome probabilities by the same matrix $\Lambda$, namely \(\mathbf{p}^{\rm exp}=\Lambda\cdot \mathbf{p}^{\rm ideal}\), due to the linearity of the Born rule.

The calibration matrix is experimentally obtained by preparing states in the computational basis and measuring them in the same basis, as reported in Sec.~\ref{init}. The entries of $\Lambda$ are then given by
\(
\Lambda_{ij}= p(i|j)
\),
namely the probability of measuring the outcome $j$ given that the preparation is in the state $\ket i$. In the ideal case, the calibration matrix should coincide with the identity matrix, i.e. one should not get outcomes different from the preparation. However, in the presence of readout noise this is not the case. 

Given the calibration matrix, one would be tempted to just invert it to obtain the ideal probabilities in the post-processing of the data, and to some extent, this can work well when the objective of the experiment is the expectation value of an observable ~\cite{Bravyi_Sheldon_Kandala_Mckay_Gambetta_2021}. However, this is considered bad practice when the actual focus of the analysis is the vector of outcomes probabilities, such as our case. The reason is that the inverse of a stochastic matrix is not positive in general; this means that applying this inverse matrix to the probability data may yield negative, non-physical probability vectors. To avoid this, we employ another post-processing, least-squares measuring filter method. This consists in computing the probabilities that minimize the distance between the measured ones given the noise matrix. The vectors obtained by this minimization are then used for the subsequent analysis. More explicitly, we compute 
\begin{equation}
	\mathbf{p}^{\rm mit}:=\argmin_{\mathbf{p}\geq 0, \sum p_i=1}\norm{\Lambda\cdot \mathbf{p}-\mathbf{p}^{\rm exp}}_2 ^2\,.
\end{equation}
This method is straightforward, but difficult to scale to more than a few qubits, since in the case of correlated noise we need to store the entire $2^N\times 2^N$ calibration matrix in memory and perform a minimization procedure over a $(2^N\!\!-\!1)$-dimensional space of probabilities. However, since the number of qubits involved in our experiments is either one or two, this method is easily implementable and quick. In case of larger qubit number, other methods may be implemented~\cite{Kandala2019,Song2019,Bravyi_Sheldon_Kandala_Mckay_Gambetta_2021,Alistair2021,VanDenBerg2022,Maciejewski_Zimboras_Oszmaniec_2020,Cosco_Plastina_Gullo_2025}.

\textit{\textbf{Depolarizing noise}} On the prediction side, we account for the presence of depolarizing noise on the $\cnot$ gate (or rather, the $\rm CZ$ since it is the native gate of our processor), which explains the lower gate fidelity compared to the single qubit ones, as discussed in Section “Experimental Protocol" in the main text. The action of the depolarizing noise channel on a state $\rho$ is described as follows~\cite{Nielsen2010}:

\begin{equation}\label{app:dep_def}
	\mathcal{E}_{\rm dep}(\rho):=p_{\rm dep}\,\rho+ (1-p_{\rm dep})\, \frac{\Id}{d}\, ,
\end{equation}
which means that with a probability $1-p_{\rm dep}$ the initial state is mapped to the completely mixed state, whereas with probability $p_{\rm dep}$ the state remains unchanged. We used the respective parameters of depolarizing noise, which correspond exactly to the survival probability $p_1$ of the Interleaved Randomized Benchmarking (IRB)~\cite{Magesan2012} seen in Eq. S4 in \textit{Supplemental Information} for each experiment.

\begin{table*}[ht]
	\caption{\label{exp_tab} Systematic investigation of experimental SRE for single and two-qubit quantum states, for qubits D3 and C4, as a function of: the readout fidelity $F_{RO}$, the average single-qubit gates fidelity from Randomized Benchmarking $F_{RB}$, relaxation, Ramsey and Echo coherence times ($T_1$, $T_2^{*}$ and $T_2^{\rm Echo}$, respectively). For two-qubit circuits, we report an estimation of the drive and ZZ crosstalk, $C^{\rm MV}$ and $C^{\rm ZZ}$ respectively, on the investigated pair. The final columns refers measured SRE for the states after introduction of $T$ gates and in the circuits are denoted as $\vert \psi_i^{T} \rangle$.}
	\begin{tabular}{cccccccccc}
		\hline
		\hline
		\textbf{Circuit} & \textbf{Qubit} & $\mathbf{F_{RO} (\%)}$ & $\mathbf{F_{RB} (\%)}$ & $\mathbf{T_2^{\mathbf{ echo}} (\boldsymbol{\mu} s)}$ & $\mathbf{T_2^* (\boldsymbol{\mu} s)}$ & $\mathbf{T_1 (\boldsymbol{\mu} s)}$ & \textbf{state} & $\mathbf{M_2}$ ($\mathbf{ = 0}$)  & $\mathbf{M_2}$ ($\mathbf{\neq 0}$) \\
		\hline
		\multirow{4}{*}{Fig.~1(c)} & \multirow{3}{*}{D3} & \multirow{2}{*}{$96 \pm 1$} & \multirow{2}{*}{$99.75 \pm 0.01$}& \multirow{2}{*}{$34 \pm 2$} & \multirow{2}{*}{$24 \pm 1$} & \multirow{2}{*}{$30 \pm 1$} & $\vert \psi_0 \rangle $ & $0.1 \pm 0.3$ & \\
		%\hhline{|~|~|~|~|~|~|~|-|-|-|}
		&   &  & & & & & \multirow{3}{*}{$\vert \psi_1\rangle$} & $0.0 \pm 0.3$& $0.43 \pm 0.12$ \\
		%\hhline{|~|~|-|-|-|-|-|~|-|-|}
		& \multirow{3}{*}{C4}  & $97 \pm 1$ & $99.910 \pm 0.002$ & $38 \pm 1$  & $34 \pm 4 $ & $30.0 \pm 0.4$ & & $0.0 \pm 0.3$ & $0.42 \pm 0.12$ \\
		%\hhline{|~|-|-|-|-|-|-|~|-|-|}
		& & $94 \pm 1$ & $ 99.925 \pm 0.002$ & $ 42 \pm 1$ & $21 \pm 1$ & $ 36 \pm 2$ &  & $0.0 \pm 0.3$ & $0.39 \pm 0.14$ \\
		\hline
		\multirow{2}{*}{Fig.~1(d)} & D3  & $97 \pm 1$ &  $99.910 \pm 0.002$ & $ 38 \pm 1$& $34 \pm 4$ & $ 30.0 \pm0.4$& \multirow{2}{*}{$|\psi_2\rangle$} & $0.0 \pm 0.3$ & $0.43 \pm 0.12$\\
		%\hhline{|~|-|-|-|-|-|-|~|-|-|}
		& C4  & $94 \pm 1$ & $99.925 \pm 0.002$ & $42 \pm 1$& $21 \pm 1$ & $36 \pm 2$ &  & $0.0 \pm 0.3$ & $0.40 \pm 0.13$\\
		\hline \hline
		\multicolumn{2}{l}{} & \multicolumn{1}{c}{\textbf{$\mathbf{F_{RO} (\%)}$}} & $\mathbf{F_{RB} (\%)}$ & $\mathbf{F_{RB}^{CZ} (\%)}$ & $\mathbf{C^{MW} (\%)}$ & $\mathbf{C^{ZZ} (kHz)}$ \\
		\hline
		\multirow{2}{*}{Fig.~1(e)}& \multirow{4}{*}{\shortstack{D3 \\ C4}}  & \multirow{2}{*}{$92 \pm 1$} & \multirow{2}{*}{$99.75 \pm 0.01$} & \multirow{4}{*}{$98 \pm 2$} & \multirow{2}{*}{$\sim 0.12$} & \multirow{4}{*}{$\sim 100$}& \multirow{2}{*}{$\vert \psi_3\rangle$} & \multirow{2}{*}{$0.1 \pm 0.1$} &\multirow{2}{*}{$0.85 \pm 0.05$} \\
		& &  &  &  & & & &  & \\
		%\hhline{|-|~|-|-|~|-|~|-|-|-|}
		\multirow{2}{*}{Fig.~1(f)} & & \multirow{2}{*}{$92 \pm 1$} & \multirow{2}{*}{$99.905 \pm 0.003$} &  & \multirow{2}{*}{$\sim 1.57$} &  & \multirow{2}{*}{$\vert \psi_4\rangle$} & \multirow{2}{*}{$0.2 \pm 0.1$} & \\
		&  &  &  &  & & &  &  & \\
		\hline
	\end{tabular}
\end{table*}

\section{Experimental setup}
\label{experimental_setup}

\begin{figure*}[t]
	\centering
	\includegraphics[width=0.7\textwidth]{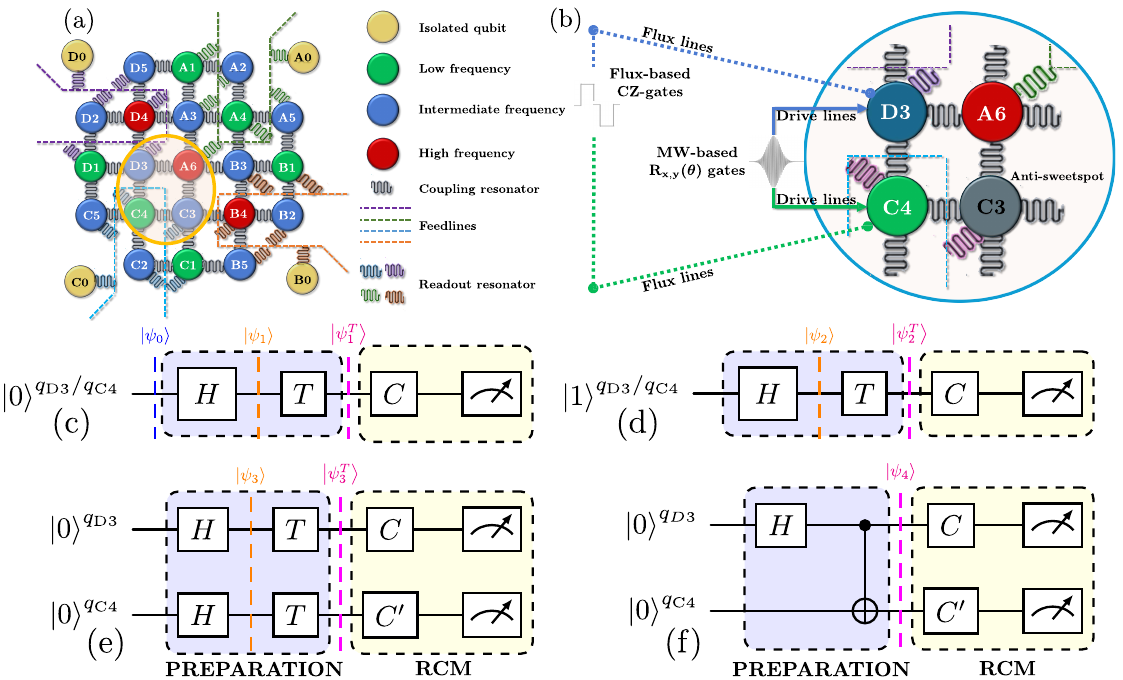}
	\caption{Superconducting Quantum Processing Unit and Benchmarking Circuits.
		(\textbf{a}, \textbf{b}) Processor schematics. In (a), overview of the superconducting quantum processor. In (b), focus on the investigated pair (D3–C4), coupled with a high-frequency qubit (A6) parked at its sweet spot and a low-frequency qubit (C3) at the anti-sweet spot (in black). 
		(\textbf{c}–\textbf{f}) Benchmarking circuits. The colored quantum states correspond to the injection (purple) and non-injection (orange) of non-Clifford $T$ gates. The measurement block represents the RCM protocol.\vspace{-0.5cm}}
	\label{processor_and_circuits}
\end{figure*}

\begin{figure}[h!]
	\begin{center}
		\includegraphics[width=1\textwidth]{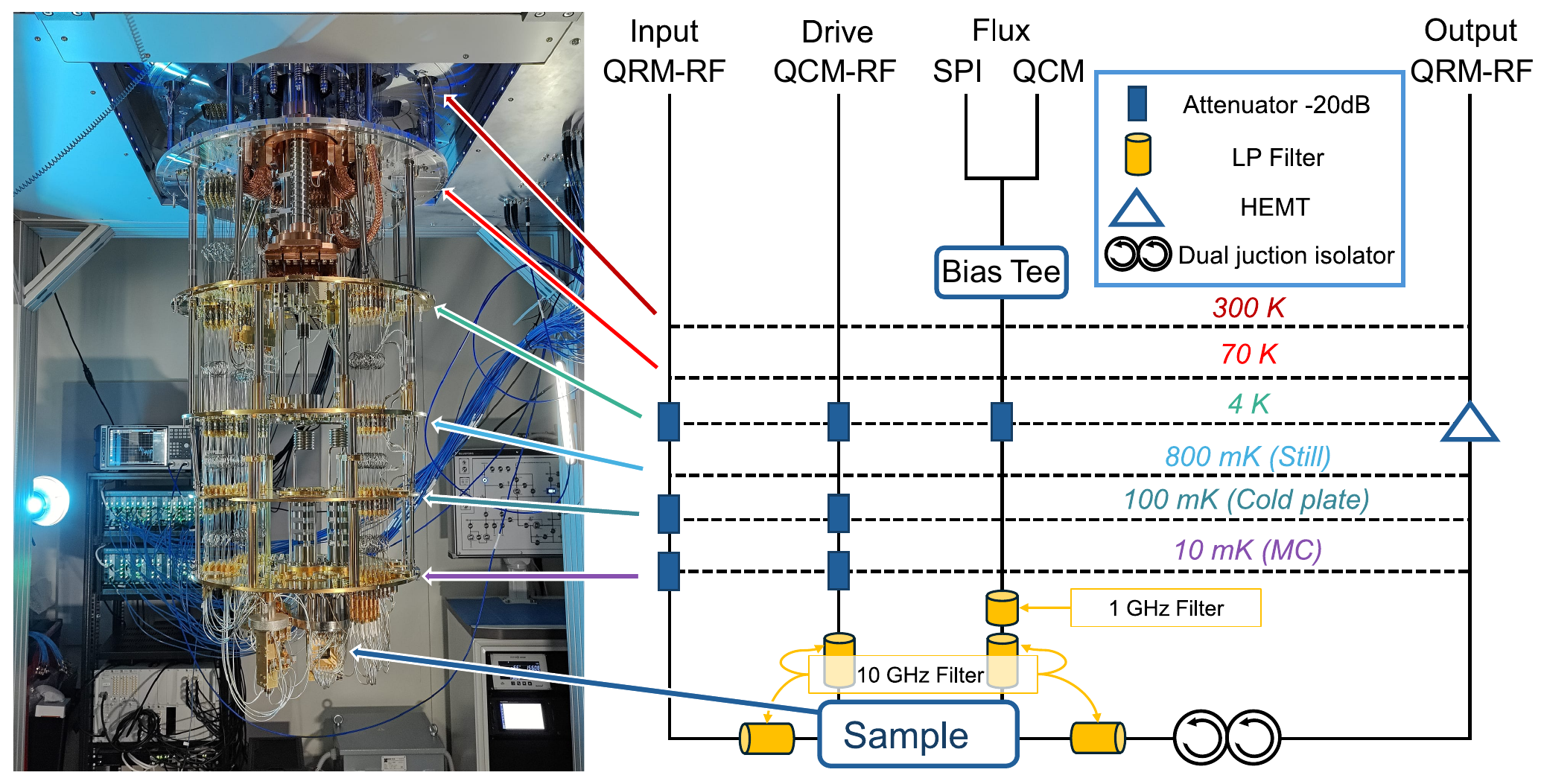}
	\end{center}
	\caption{Cryogenic setup scheme, including the attenuation scheme for the input, drive and flux lines. On the output line, there are a dual juction isolator and an HEMT amplifier. Each line has a low-pass filter.}
	\label{setupcryo}
\end{figure}

The fridge employed in this work is a dry dilution refrigerator BlueforsXLS1000SL, with minimal base temperature of  $7\;mK$. 
It is equipped with microwave coaxial cables, which allow
operation within the microwave frequency range (DC- $18 GHz$, nominally). The cryostat features four types of cryogenic lines: input and output lines for the readout, drive lines for qubit control, and flux lines for frequency tuning with either static or pulsed external magnetic flux. Feedline and drive lines feature a nominal attenuation of $-60dB$, which adds to $-10dB$ attenuation of the line itself (stainless steel) and infrared low-pass filters with a cutoff of about $10GHz$. The flux lines are equipped with $-20dB$ nominal attenuation, anchored at $4K$ to prevent Joule heating, and two infrared low-pass filters in series of $10 GHz$ and $1 GHz$. Flux lines are in stainless steel from room temperature to the 4K plate, and in niobium titanium below $4K$. This allows for reducing heat dissipation when delivering static currents up to $40mA$, used to generate magnetic flux on chip. The output lines, also made of niobium titanium below $4K$ and in copper-nickel at highest temperatures, are not attenuated, and include a $10 GHz$ low-pass filter and a Low-Noise Factory LNF-ISC double junction isolator at the mixing chamber (MXC), so that signals from room-temperature towards the sample are attenuated nominally by a total of $40\;dB$. Since the output signals of the qubits are single-photon signals, amplifiers are required. In our system, there are two amplification stages. There is a High Electron Mobility Transistor (HEMT) with nominal amplification of $40 dB$ on the 4K plate, and three amplifiers at room temperature with a nominal $16 dB$ amplification each. The cryogenic setup is shown in Figure \ref{setupcryo}.
Finally, to provide combined static and pulsed magnetic flux through the flux lines, we employ bias-tees at room temperature, where the DC signal is generated by Qblox S4g modules (installed in a Qblox SPI-rack) and the RF signal is generated by a Qubit Control Module (QCM). This generates output signals up to 400 $MHz$, and are installed in a Qblox Cluster, a room-temperature microwave modular electronics fully interfaceable with Python packages, thanks to the open-source package Quantify~\cite{Quantify}. 
The modules used for the measurements of this work are:
\begin{itemize}
	\item Qubit Readout Module RF (QRM-RF), which allows generation of output multi-tone signals for multiplexed readout up to $18.5\;GHz$, and to demodulate the readout signals over a $400\;MHz$ bandwidth;
	\item Qubit Control Module RF (QCM-RF), which allows output signals up to $18.5\;GHz$;
	\item Qubit Control Module (QCM), which generates pulses in the baseband regime up to $400\;MHz$.
\end{itemize}

\begin{figure}[h!]
	\begin{center}
		\subfloat[][]{\includegraphics[width=0.8\textwidth]{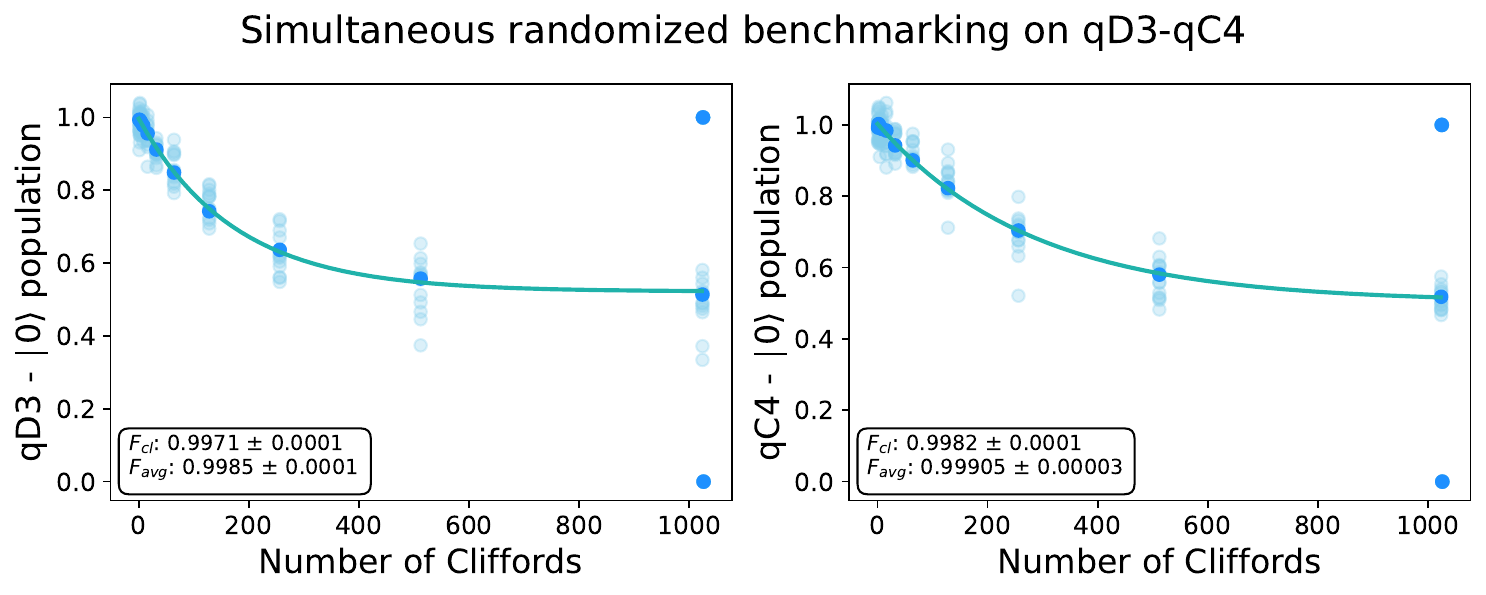}}\\
		\subfloat[][]{\includegraphics[width=0.8\textwidth]{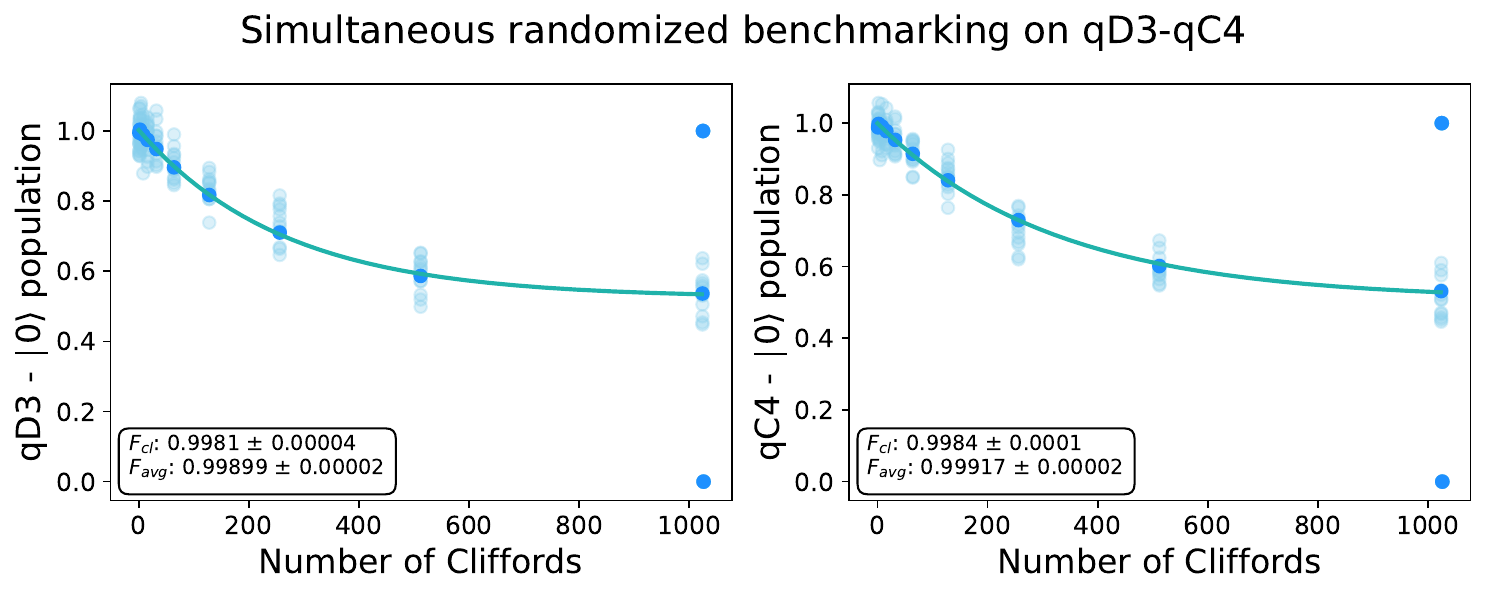}}\\
		\subfloat[][]{\includegraphics[width=0.8\textwidth]{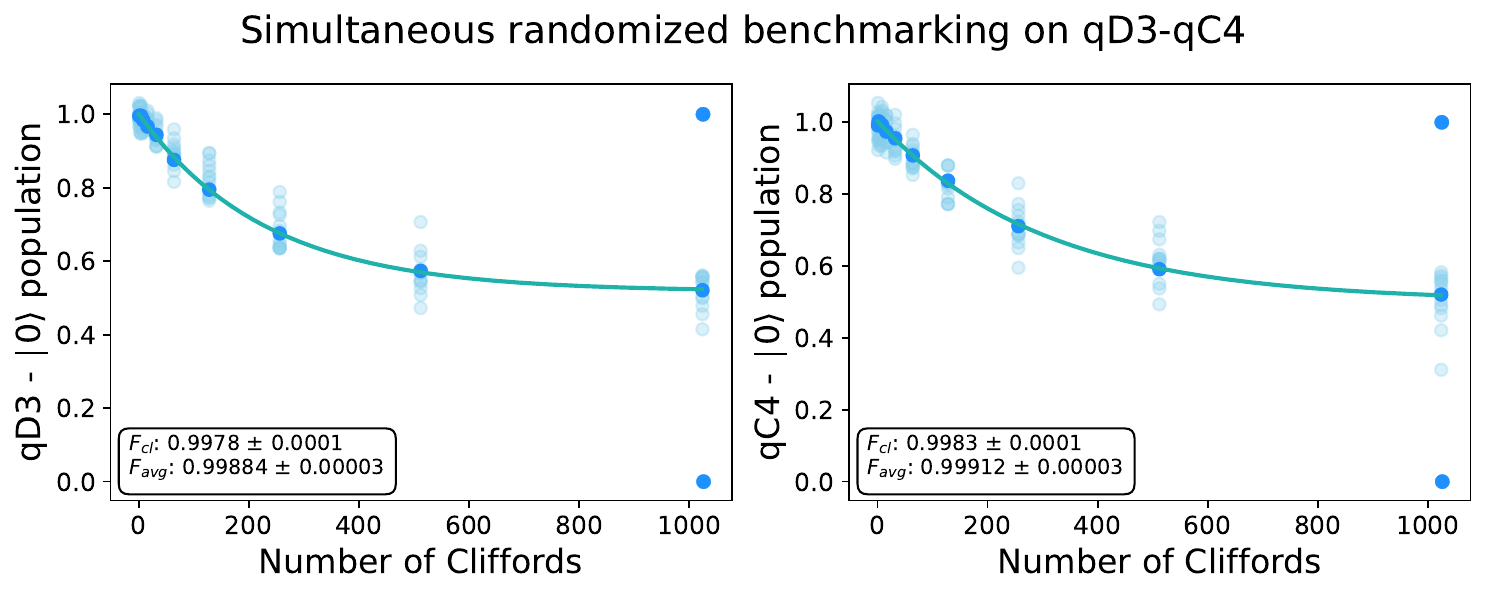}}
	\end{center}
	\caption{Simultaneous Randomized Benchmarking for single-qubit gate fidelity on qubit qD3 and qC4 for the three proposed experiment: the RB in (a) for the LM state experiment, in (b) for the M state, and in (c) for the NLM state. Blue scatter data corresponds to the population of the $\ket{0}$ state as a function of the number of Clifford gates in the randomized benchmarking sequence, averaged over $50$ random seeds. The solid line represents the fit function (Eq.\eqref{exp_dec}) used for the estimation of the single-qubit randomized benchmarking average gate fidelity.}
	\label{rb_sim}
\end{figure}
\begin{figure}[h!]
	\begin{center}
		\subfloat[][]{\includegraphics[width=0.5\textwidth]{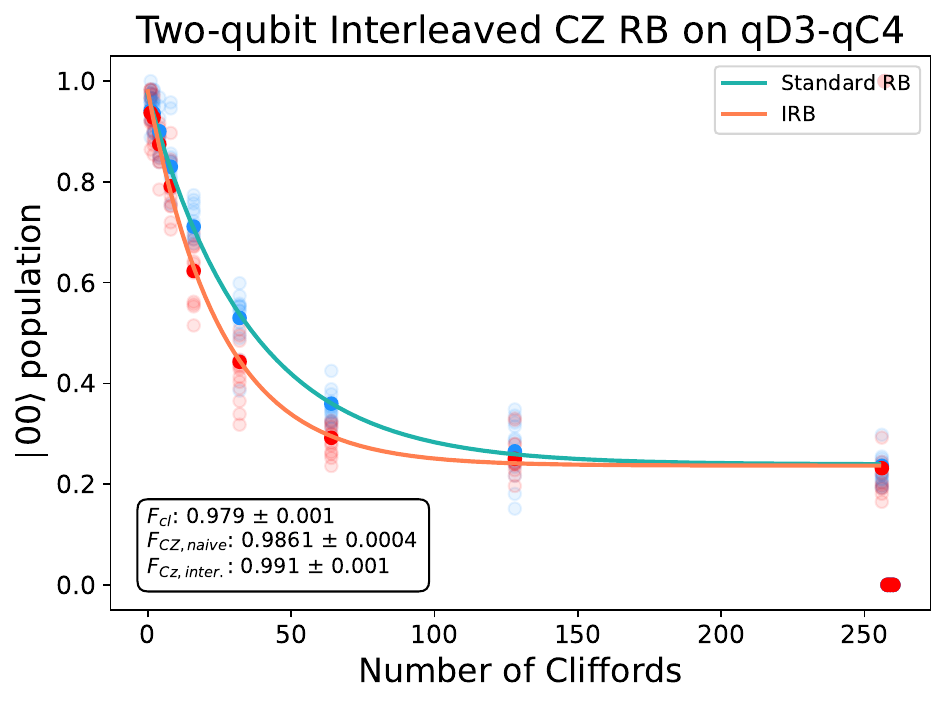}}
		\subfloat[][]{\includegraphics[width=0.5\textwidth]{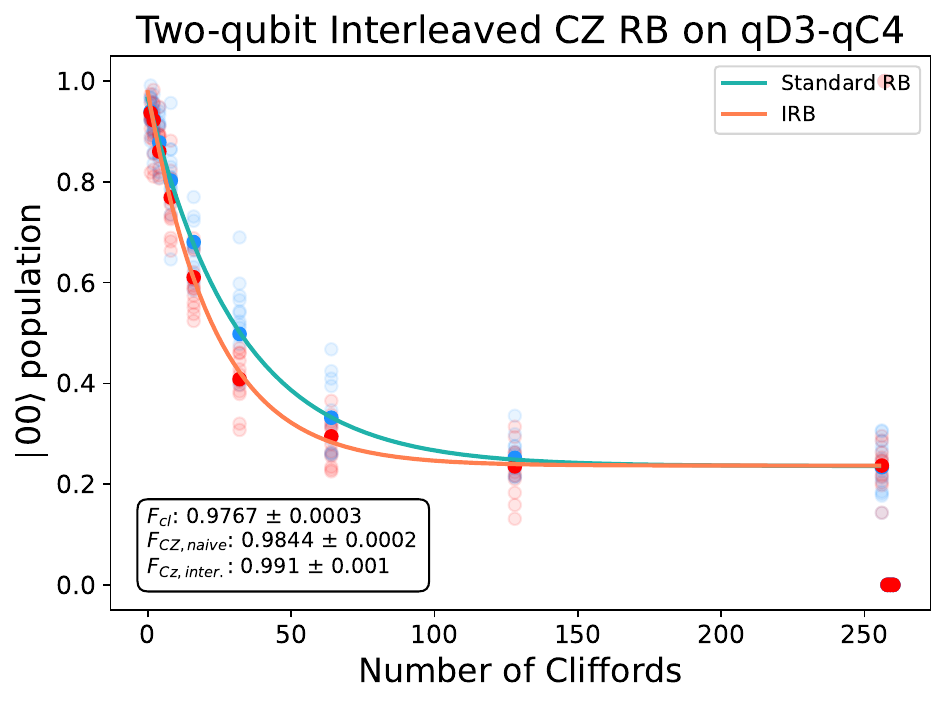}}\\
		\subfloat[][]{\includegraphics[width=0.5\textwidth]{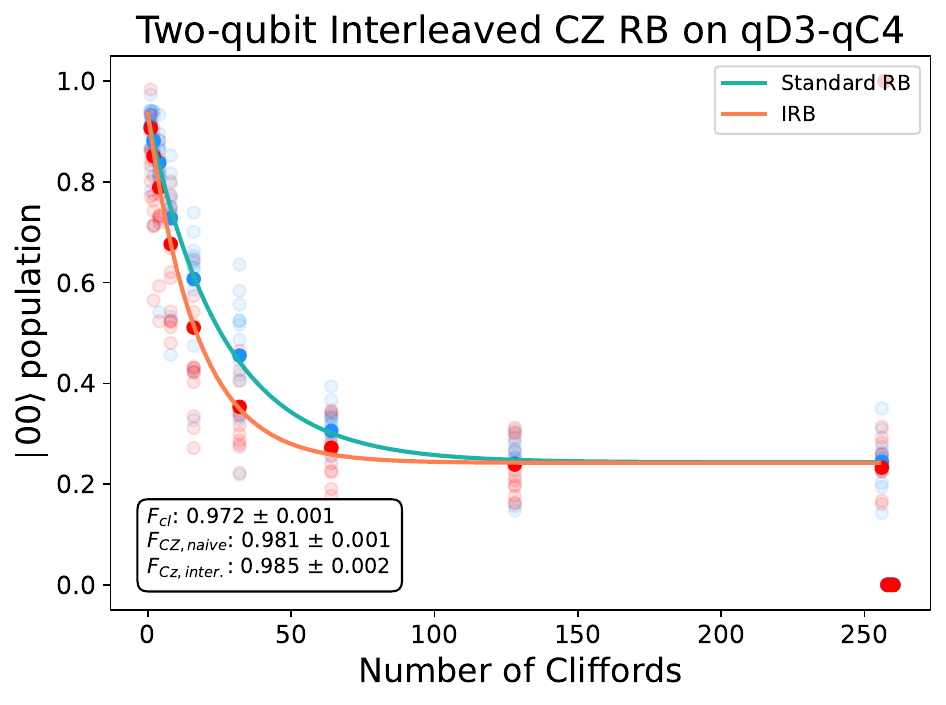}}
	\end{center}
	\caption{Interleaved Randomized Benchmarking for the CZ gate on qubit qD3 and qC4 for the three proposed experiment: the CZ-IRB in (a) for the LM state experiment, in (b) for the M state, and in (c) for the NLM state. The scatter data correspond to the population of the $\ket{00}$ state as a function of the number of Clifford gates in the randomized benchmarking sequence, averaged over $50$ random seeds. The solid line represents the fit function (Eq.\eqref{exp_dec}) used for the estimation of survival probabilities. Specifically, the blue line refers to the standard simultaneous RB, while the orange one refers to the IRB, from which it is possible to estimate $p_0$ and $p_1$, respectively. From the two survival probabilities, the fidelity of the CZ can be obtain according to Eq.\eqref{cz_eq}.}
	\label{rb_int}
\end{figure}

\section{Benchmarking protocols and experiments}
\label{App.Benchmark}

\subsection{Single and two-qubit gates optimization protocols}

Before the experimental measurements of magic on the investigated D3-C4 pair, we have performed: i) single- and two-qubits gate calibrations~\cite{Lucero2008,Reed}, ii) simultaneous readout calibration. In detail, the experimental single-qubit gate set includes rotations $R_{xy}(\theta)$, where the rotation angle $\theta$ is controlled by changing the amplitude of drive DRAG (Derivative Reduction Adiabatic Gate) pulses~\cite{Krantz2019,Babu2021,Werninghaus2021} with a fixed duration of $60\;ns$. After a rough estimation of the amplitude required to perform a transition from the ground to the excited state of the qubits ($\pi$-pulse) with Rabi oscillations experiments~\cite{Krantz2019}, we have optimized the pulses frequency by using Ramsey oscillations experiments~\cite{Krantz2019}, the shape through Motzoi protocol~\cite{Motzoi2009} and we performed a fine tuning of the $\pi$-pulse amplitude through the flipping protocol. This sequence of optimization steps is finally tested through simultaneous All-XY experiments, able to recover error syndromes in X-Y gates (e.g., detuning, amplitude and DRAG errors), and eventually allowing us to correct for specific errors when they occur~\cite{Reed}. 

As for the two-qubit gates calibration, we have focused on CZ gates implemented through flux-pulses able to set on resonance the high-frequency qubit (D3) with the low-frequency qubit (C4). We used the cryoscope technique to correct for flux pulses distortions~\cite{Rol2020}, and CZ-Chevron experiments employing unipolar flux pulses to get a rough estimation of the amplitude and the duration of the CZ flux pulses~\cite{Krantz2019,Negirnac2021}. Then, we have used Sudden-Net Zero (SNZ) flux pulses, whose parameters have been optimized through conditional oscillations experiments~\cite{Rol2019} (Sec.~\ref{App.Benchmark}). By measuring landscapes of the conditional two-qubit phase and the leakage as a function of the amplitude and the shape of the SNZ pulses, as well as the flux sweet spot of the qubits, we aimed at the minimum leakage and a conditional phase as close as possible to the nominal $180^\circ$ conditional two-qubit phase~\cite{Negirnac2021}. Finally, we have also corrected for the single-qubit phases to enhance the fidelity of the CZ gates~\cite{Negirnac2021}. 

The readout calibration procedure, instead, involved Single-Shot Readout (SSRO) experiments~\cite{Mallet2009} as a function of the power and the frequency of the readout tone, as well as the duration of the readout signal and the integration time. Single-shot readout is performed suddenly before any quantum circuit implemented, and is used to identify the discrimination threshold for the computational basis state and the calculation of the count vectors and the readout state probability~\cite{Stasino2025}. 
Finally, besides standard coherence times benchmarking, e.g. relaxation, Ramsey and Hahn-Echo coherence times ($T_1$, $T_2^{*}$ and $T_2$)~\cite{Krantz2019}, simultaneous single-qubit Randomized Benchmarking (RB)~\cite{Gambetta2012} and CZ Interleaved two-qubits RB (IRB)~\cite{Magesan2012,Corcoles2013} were used to estimate the average Clifford gate fidelities for the two qubits and the average CZ gate fidelity (App.~\ref{App.Benchmark}), while SSRO experiments have been used to estimate the readout fidelity~\cite{Stasino2025}.

In the context of RCM, $C=\bigotimes_{i=1}^N C_{i}$ represents the tensor product of $N$ single-qubit Clifford unitaries $C_i$, each randomly sampled from the ensemble of unitaries, which includes $X$, $Y$, $Z$, $H$, and $S$ gates, plus the trivial identity gate $I$, and all 24 of their possible combinations. While $X$, $Y$ and $Z$ gates are directly related to microwave/flux pulses\cite{Krantz2019}, $H$ and $S$ gates are decomposed in terms of $R_x$, $R_y$, and $R_z$ gates. For example, $H$ is decomposed in terms of $R_y(\pi/2)R_z(\pi)$ rotations, while $S$ is implemented as a $R_z(\pi/2)$ pulse, where $R_{y} (\theta)\!\! \equiv \!\! R_{x y} \left(\theta, \frac{\pi}{2}\right)\! $,  $R_z(\theta) = e^{-i \frac{\theta}{2} Z}$. The $T$ is implemented as $R_{z} (\pi/4)$. $\rm CNOT$ is implemented as $ \mathrm{CNOT}_{1 \to 2} = (I \otimes H) \cdot \mathrm{CZ}_{1 \to 2} \cdot (I \otimes H)$.

%Single- and two-qubits Gate Set Tomographies (GST) and Bell states Quantum State Tomographies (QST) have also been used to quantify the fidelity of specific gates and states employed in the quantum circuits implemented in this work.  

\subsection{Fidelity estimation}

Randomized Benchmarking (RB) is a class of methods used to evaluate the fidelity of applied gates and is based on the implementation of sequences of random gate operations\cite{Hashim2025,Helsen2022RBGeneral,Proctor2022}.
The advantages of RB protocols are their efficient scalability with the number of qubits $n$ and the fact that they account for State Preparation and Measurement (SPAM) errors.
In our protocol, the RB has been implemented over the Clifford gate set.
Clifford based RB has several variants. Specifically, we focused on the Standard RB and the Interleaved RB (IRB).

\subsubsection{Single qubit Randomized Benchmaking}
Standard Clifford RB is performed by initializing the qubit in the $\ket{0}$ state, followed by the application of a sequence of $N_{cl}$ random Clifford gates, and then the inverse of all applied gates. Ideally, this sequence acts as the identity gate. However, due to gate errors, the final state will differ from the ideal one. Finally, we can thus extract the fidelity of the gates by comparing how close the final state is to the initial one. The survival probability $p$ can be obtained by fitting the population of the $\ket{0}$ state as a function of the number of Clifford, using the exponential decay\cite{Epstein2014}:
\begin{equation}
	F_{\ket{0}}(N_{ \rm cl}) = A \cdot p^{N_{ \rm cl}} + B,
	\label{exp_dec}
\end{equation}
where $N_{\rm cl}$ is the number of Clifford gates in the sequence. 
Given $d = 2^n$ as the dimensionality of the system for $n$ qubits, the average gate fidelity can be estimated from the survival probability as:
\begin{equation}
	F_{\rm avg.gate} = (F_{\rm cl})^{\frac{1}{1.875}},
\end{equation}
where
\begin{equation}
	F_{\rm cl} = 1-\left(\frac{d-1}{d}-(1-p)\right).
\end{equation}
This protocol was performed simultaneously on both qubits, qD3 and qC4. The single-qubit gate for each qubit was chosen randomly and independently from the single-qubit Clifford group, i.e., the protocol required applying the product of the single-qubit Clifford unitaries to the two-qubit register.
An example of the protocol applied simultaneously on qubits D3 and C4 is shown in Fig.\ref{rb_sim}.

\subsubsection{CZ - Interleaved Randomized Benchmarking}\label{irbcz}
\begin{figure}[h!]
	\begin{center}
		\includegraphics[width=0.7\textwidth]{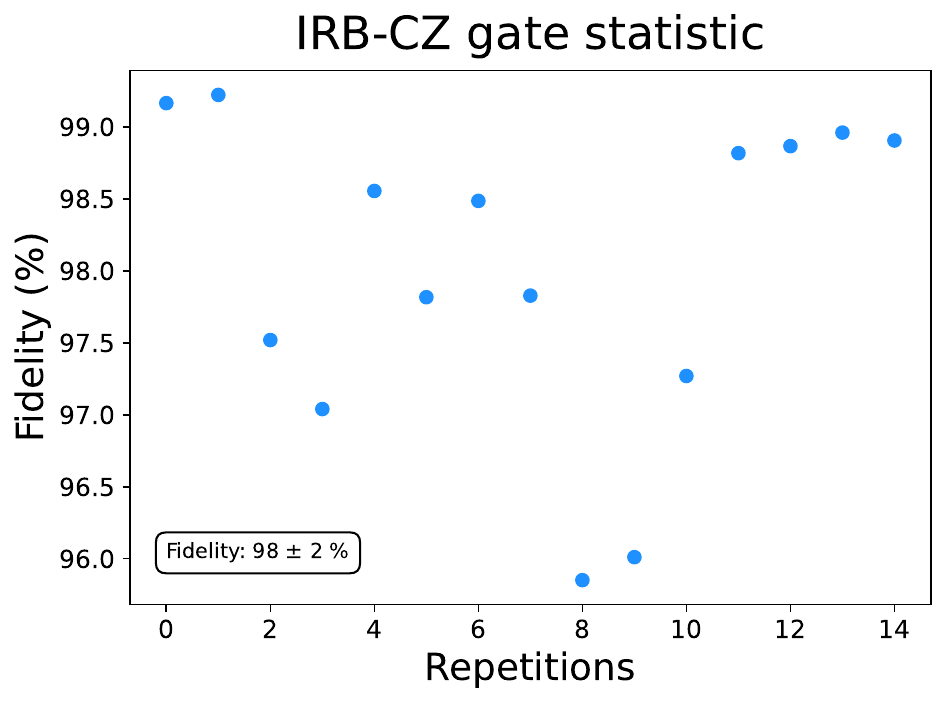}
	\end{center}
	\caption{Statistical measurements of the interleaved CZ gate fidelity.}
	\label{IRB_CZ_stat}
\end{figure}

The Interleaved Randomized Benchmarking (IRB) is a scalable experimental protocol for estimating the average error of individual quantum computational gates. The protocol involves interleaving random Clifford gates with the gate of interest, i.e., the CZ gate in our case\cite{Magesan2012}.
This protocol is composed of three steps\cite{Magesan2012}:
\begin{itemize}
	\item perform standard randomized benchmarking (RB) by choosing $K$ sequences of random gates, where the first $m$ gates in each sequence are selected randomly from the Clifford group, and the $(m+1)$-th gate is the inverse of the composition of the first $m$ gates. By fitting the exponential decay, we can obtain the survival probability $p_0$ (Eq.\eqref{exp_dec});
	\item choose K sequences of Clifford elements where the first Clifford, in each sequence, is chosen randomly for the Clifford group, while the second is always the CZ gate, and alternate between random Clifford and the CZ gate up to $m$-th random gate. The $(m+1)$-th gate is the inverse of the composition of the first $m$ gates and the $m$ CZ gates. As well as for the standard RB, we can use the exponential decay fit to estimate the survival probability $p_1$;
	\item estimate the fidelity for the CZ gate as follow:

	\begin{equation}
		F_{ \rm CZ,int} = 1- \Big(\frac{d-1}{d} \cdot \Big(1-\frac{p_1}{p_0}\Big)\Big)
		\label{cz_eq}
	\end{equation}
\end{itemize}
An example of the CZ-IRB performed on D3 and C4 is shown in Fig.\ref{rb_int}. Moreover, we performed statistical measurements of the interleaved CZ gate fidelity, obtaining an average fidelity of $(98 \pm 2)\%$. The results are shown in Fig.\ref{IRB_CZ_stat}.
\begin{figure}[h!]
	\begin{center}
		\subfloat[][]{\includegraphics[width=0.485\textwidth]{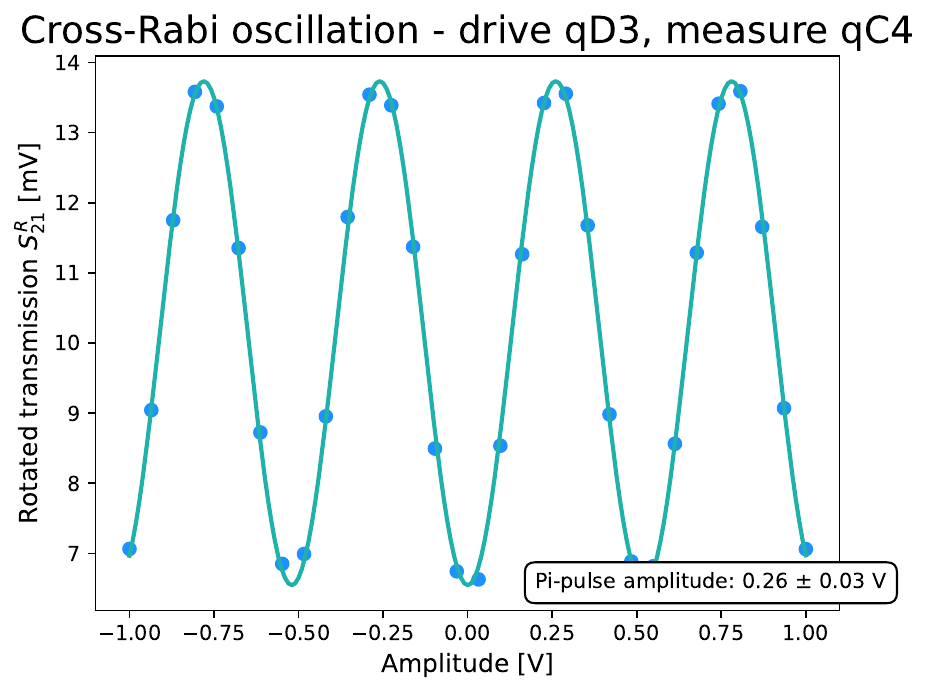}}
		\subfloat[][]{\includegraphics[width=0.5\textwidth]{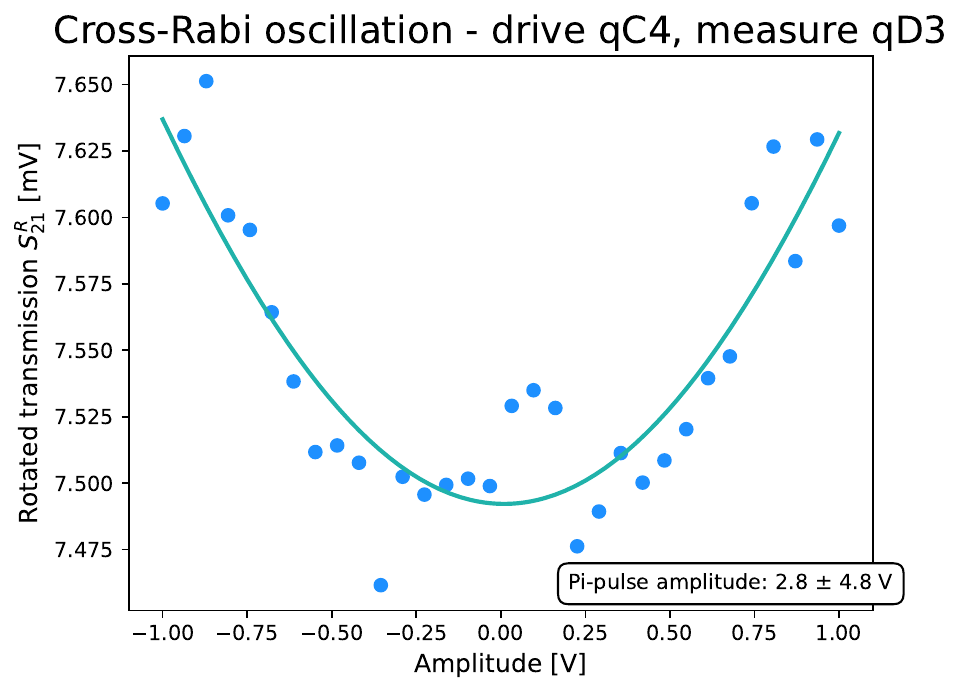}}\\
		\subfloat[][]{\includegraphics[width=0.6\textwidth]{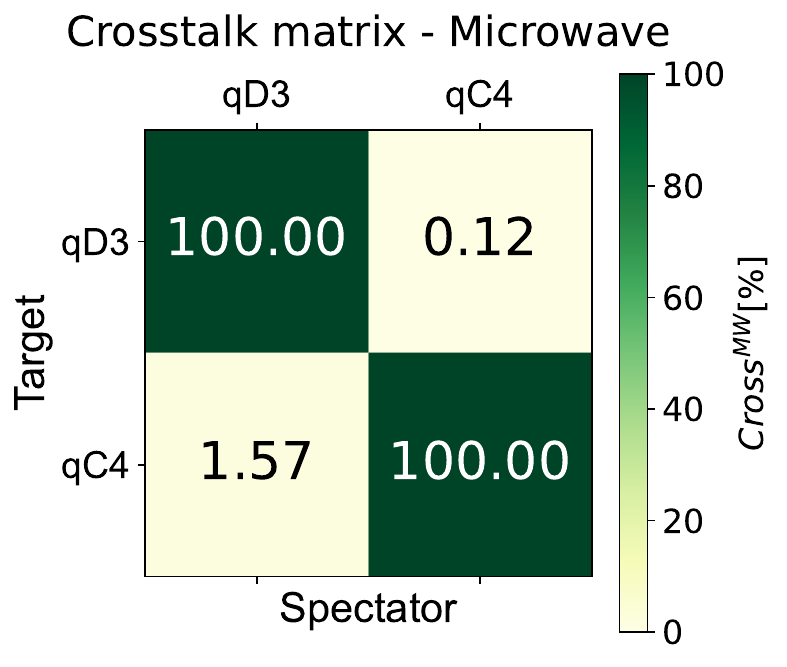}}
	\end{center}
	\caption{Measurement of the MW crosstalk between qD3 and qC4. In (a) and (b), the results of the Rabi protocol performed using qD3 as the driven qubit and qC4 as the measured one, and vice versa, respectively. In (c), the MW crosstalk matrix for qD3 and qC4. The x- and y-axis indicates the measured qubit and the driven one. The colorbar represents the percentage of MW crosstalk.}
	\label{mw_crosstalk}
\end{figure}

\subsection{Crosstalk measurements}
Crosstalk phenomena refer to interactions between qubits that can introduce noise into the system, leading to errors in algorithm implementation. Therefore, as a preliminary step, it is necessary to study the crosstalk affecting the device. There are various types of crosstalk, but in this work, we specifically focus on microwave crosstalk and ZZ crosstalk.

\subsubsection{Microwave crosstalk}
To determine the microwave crosstalk strength from one qubit to another, we send a microwave pulse to the driveline of one qubit and measure its effect on the other qubit, referred to as the 'driven' and 'measured' qubit, respectively. Specifically, we perform a Rabi experiment on the driven qubit by sending a $\pi$-pulse at the frequency of the measured qubit and carry out a dispersive measurement on the latter.
By using a sinusoidal fit, we can extract the $\pi$-pulse amplitude obtained from the experiment, which allows us to estimate the strength of the microwave crosstalk coefficient. Specifically, it can be estimated as:
\begin{equation}
	C_{i\rightarrow j} = \frac{A_{j\rightarrow j}}{A_{i\rightarrow j}} \frac{t_{j\rightarrow j}}{t_{i\rightarrow j}}
\end{equation}
where $A_{i\rightarrow j}$ and $t_{i\rightarrow j}$ are the amplitude and duration of the $\pi$-pulse sent on qubit $i$ and measured on qubit $j$, respectively.
The microwave crosstalk matrix for the couple D3-C4 is shown in Fig.\ref{mw_crosstalk}.

\subsubsection{ZZ crosstalk}
\begin{figure}[h!]
	\begin{center}
		\includegraphics[width=1\textwidth]{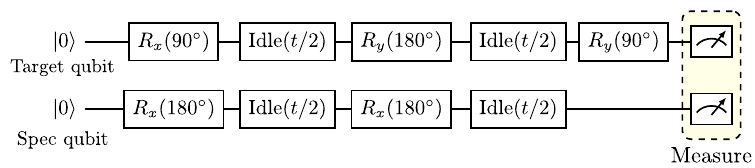}
	\end{center}
	\caption{Pulse sequence for ZZ residual coupling estimation. First, a $R_x(\pi/2)$ and $R_x(\pi)$ gates are applied on the target and the spectator qubit, respectively. The qubits are allowed to evolve freely for a variable delay $t/2$, after which a $R_y(\pi)$ and $R_x(\pi)$ rotation are applied on the target and spectator qubit, respectively. After waiting again for a variable delay $t/2$, a $R_y(\pi/2)$ pulse is applied on the target qubit. Finally, the circuit's output is measured.}
	\label{zz_res_schedule}
\end{figure}

By performing two-qubit gates, we introduce a controlled interaction between the two qubits. However, there may also be a small undesired coupling between them, which persists even when the controlled interaction is turned off. The effect of this interaction, known as residual ZZ coupling \cite{Rol2020}, is that the frequency of the $\ket{0} \rightarrow \ket{1}$ transition of one qubit shifts depending on the state of the other. In order to estimate the streght of this interaction, we can use a Ramsey-type experiment. Specifically, we perform a Ramsey sequence on the target qubit with an echo pulse at the midway point. In the meanwhile, the spectator is prepared in the ground state for half of the sequence, and in the excited state for the second half. The pulse scheme of the experiment is shown in Fig.\ref{zz_res_schedule}.
\newpage 
\begin{figure}[th!]
	\begin{center}
		\subfloat[][]{\includegraphics[width=0.85\textwidth]{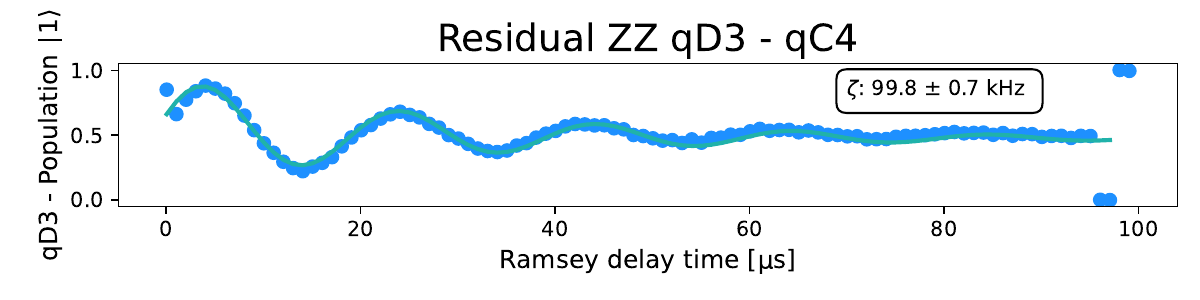}}\\
		\subfloat[][]{\includegraphics[width=0.85\textwidth]{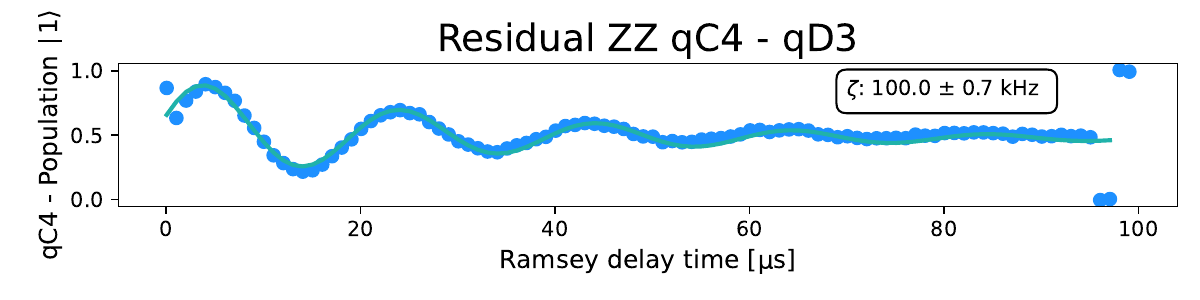}}\\
		\subfloat[][]{\includegraphics[width=0.6\textwidth]{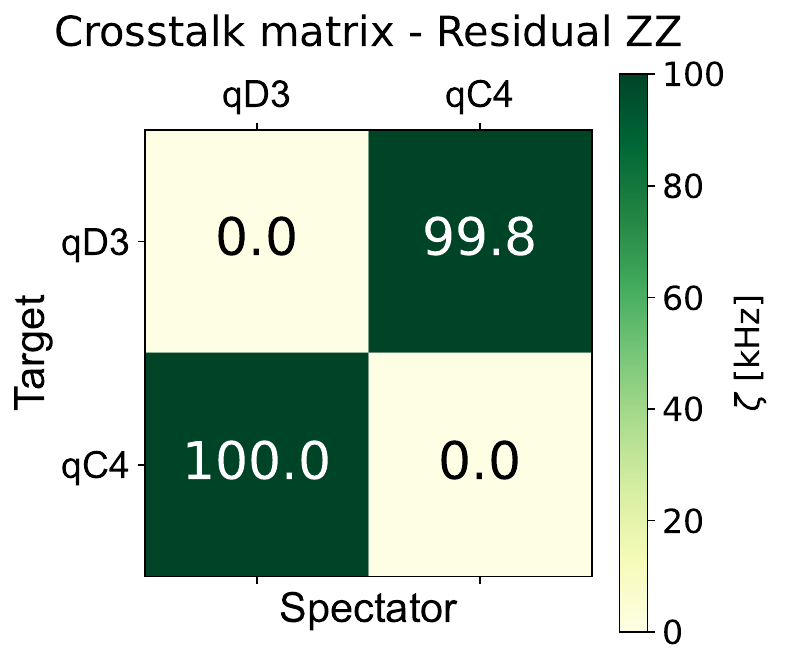}}
	\end{center}
	\caption{Measurement of the ZZ residual coupling between qD3 and qC4. In (a) and (b) the population of the $\ket{1}$ state using qD3 and qC4 as target, respectively. The solid lines represent the fit fuction used to estimate the ZZ residual coupling. In (c), the ZZ crosstalk matrix for qD3 and qC4. The x- and y-axis indicates the spectator and the target qubit, respectively. The colorbar represents the shift in frequency due to the ZZ interaction.}
	\label{zz_crosstalk}
\end{figure}
\begin{figure}[th!]
	\begin{center}
		\subfloat[][]{\includegraphics[width=0.90\textwidth]{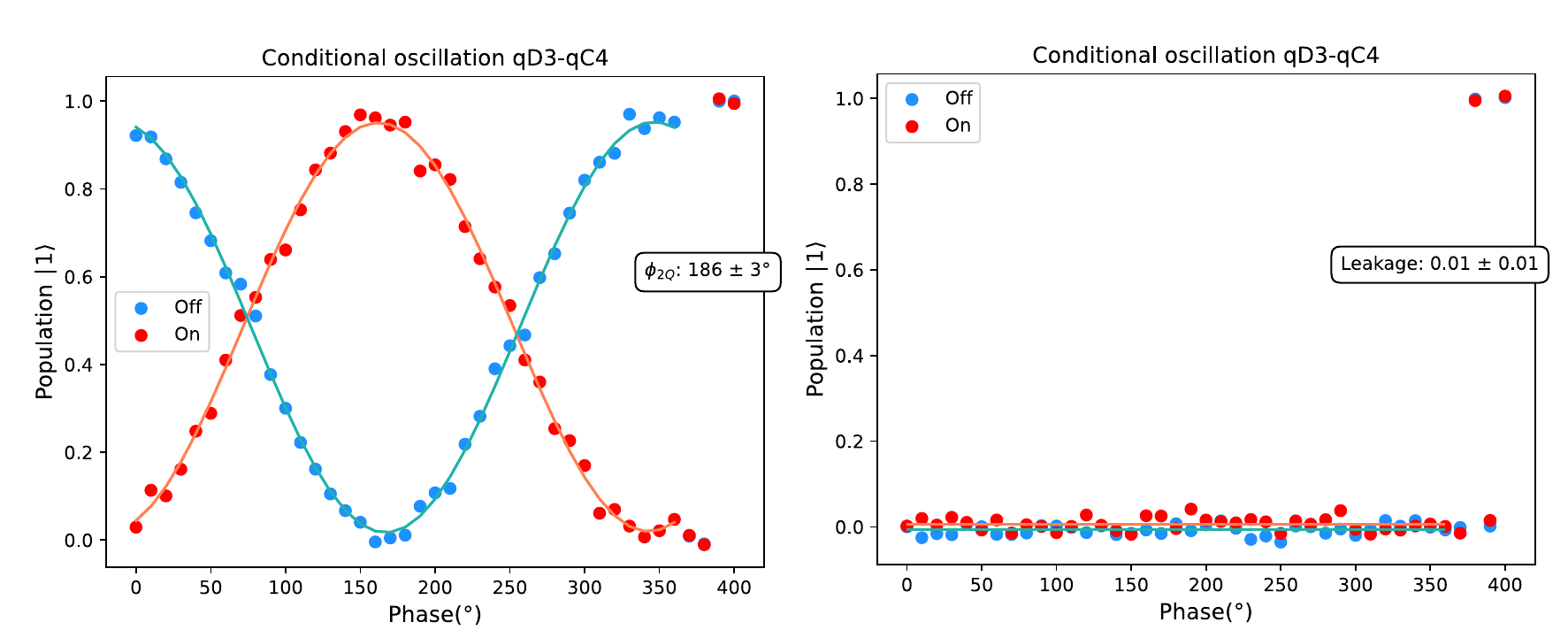}}\\
		\subfloat[][]{\includegraphics[width=0.90\textwidth]{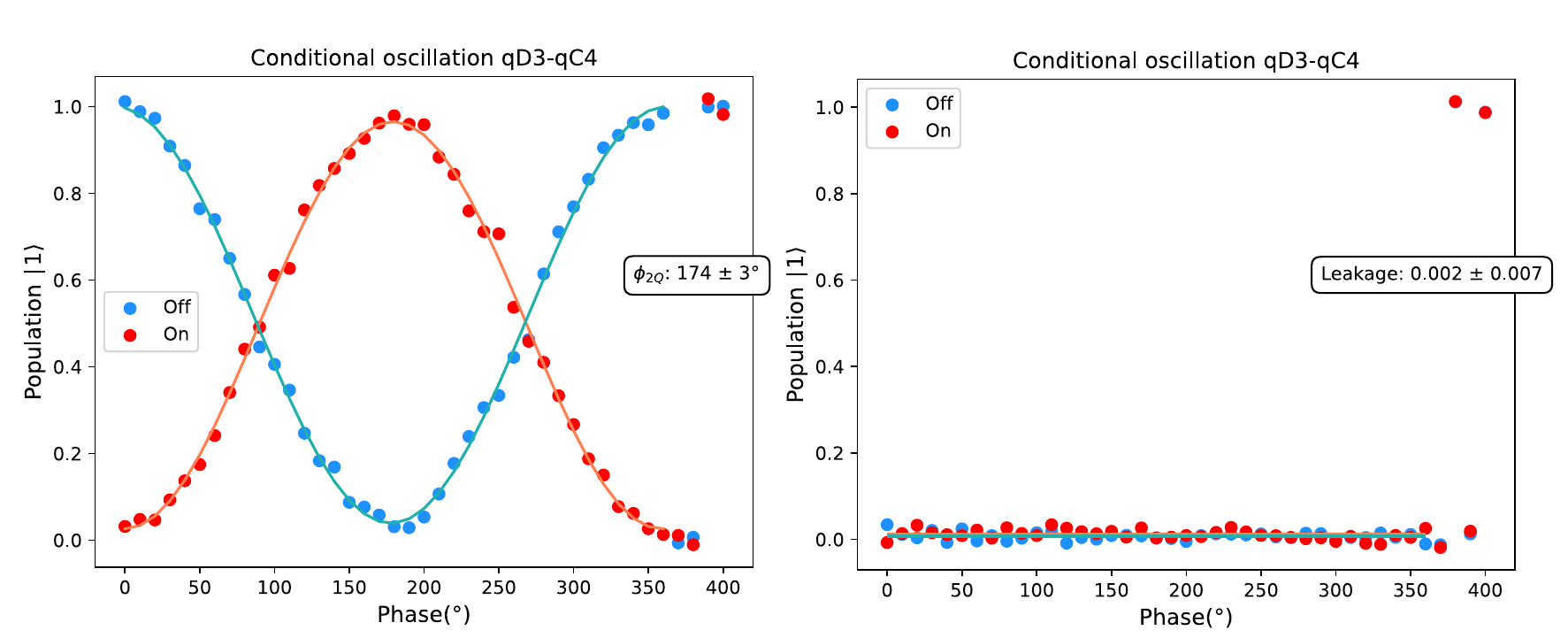}}\\
		\subfloat[][]{\includegraphics[width=0.90\textwidth]{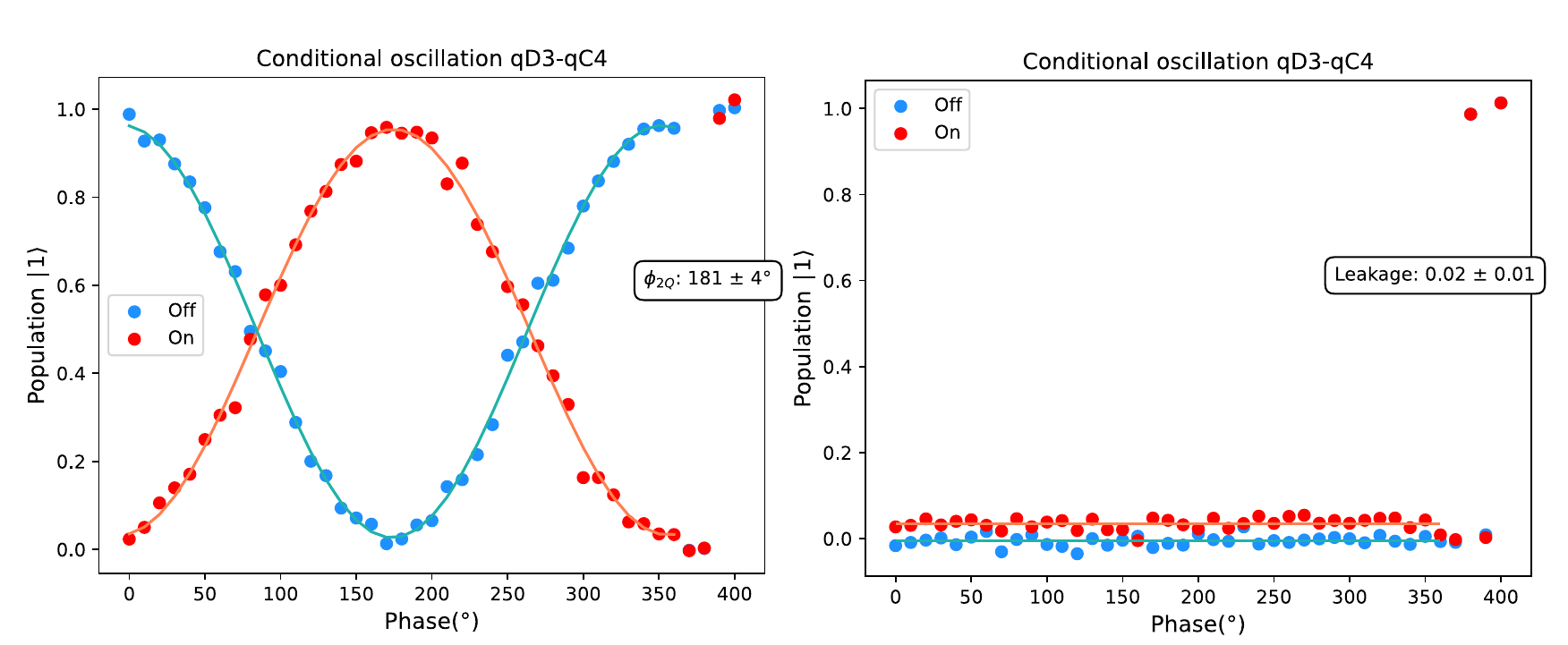}}
	\end{center}
	\caption{The measured $\ket{1}$ population as a function of the phase on qD3 on the left and on qC4 on the right for the three proposed experiment: in (a) for the LM state experiment, in (b) for the M state, and in (c) for the NLM state. The blue scatter data represent the measured values for the Off variant, while the orange ones the measured values for the On variant. The solid lines represent the fit fuction used to estimate the two-qubit phase $\theta_{2Q}$ on the left and the leakage $L$ on the right.}
	\label{CZ_cond_oscill}
\end{figure}

Due to the residual ZZ interaction, the target qubit will acquire a phase. By varying the delay in the experiment, we observe an oscillation in the target qubit population, from which it is possible to estimate the residual ZZ coupling frequency.
The ZZ-crosstalk matrix for the couple D3-C4 is shown in Fig.\ref{zz_crosstalk}.

\subsection{Conditional oscillation}

To tune up the CZ gate, it is possible to perform the conditional oscillation experiment. It can be used to measure the conditional phase $\theta_{2Q}$ acquired during an uncalibrated CZ gate, and to estimate the leakage $L$, defined as the average probability that a random computational state leaks out of the computational subspace\cite{Christopher2018}. In the conditional oscillation experiment, two variants of the same experiment are performed\cite{Rol2019}. In the first variant (Off), a $\pi/2$-pulse is applied on the target qubit, while the control qubit is left in the ground state. After that, the CZ flux pulse is applied. Finally, another $\pi/2$-pulse is applied on the target before measuring the state of both qubits simultaneously. In the second variant (On), the control qubit
is rotated into the excited state before applying the CZ gate. Then, the control qubit is pulsed back to the ground state before measuring both qubits. The difference in phase acquired by the target in the On and Off variants yields $\theta_{2Q}$, while the population difference on the control, defined as the missing fraction $m$, allows us to estimate the
leakage as $L = m/2$. In order to optimize $\theta_{2Q}$, i.e. to have it equal to $\pi$, and to minimize the leakage $L$, the flux pulse amplitude and duration are changed.
An example of the protocol performed for the CZ between D3 and C4 is shown in Fig.\ref{CZ_cond_oscill}.

\begin{figure}[h!]
	\begin{center}
		\subfloat[][]{\includegraphics[width=0.3\textwidth]{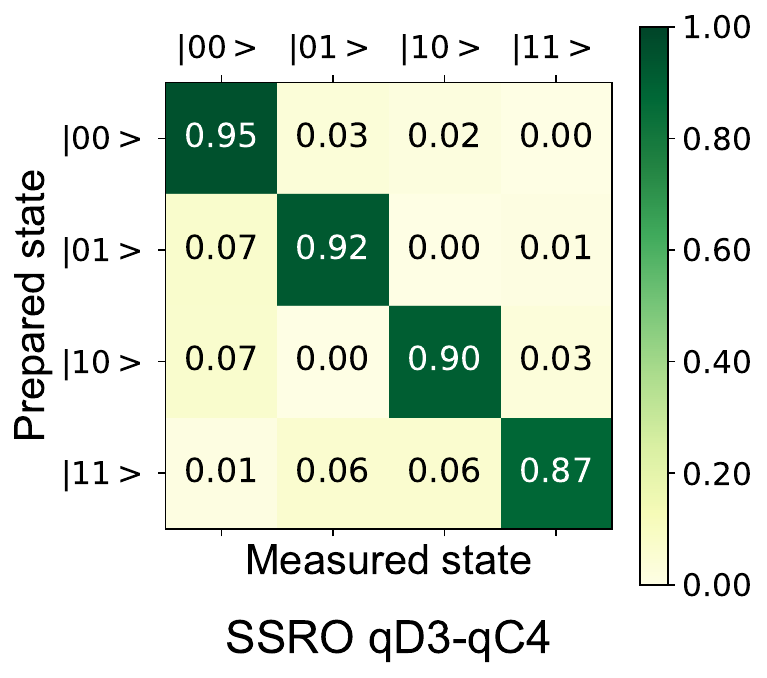}}
		\subfloat[][]{\includegraphics[width=0.3\textwidth]{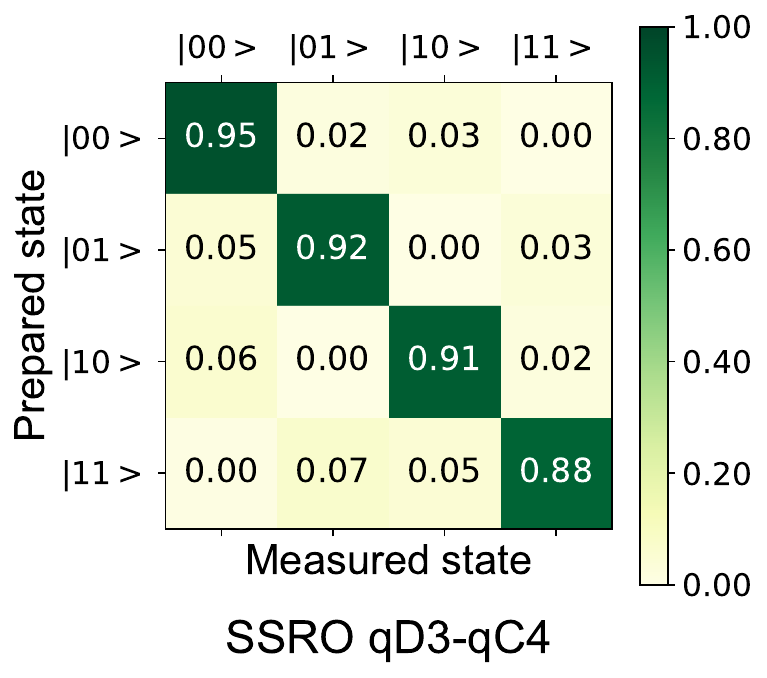}}
		\subfloat[][]{\includegraphics[width=0.3\textwidth]{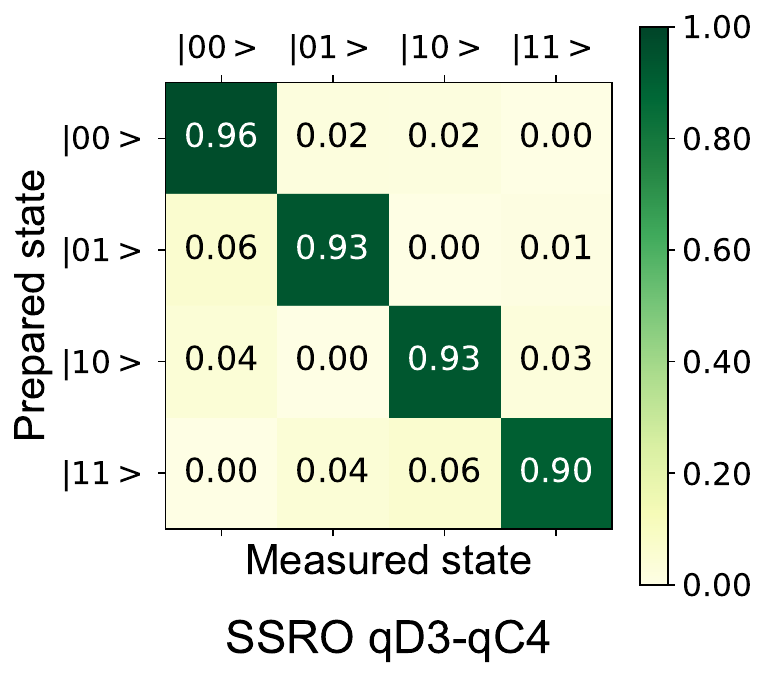}}
	\end{center}
	\caption{\textbf{Supplemental Figure - Initialization/assignment probability matrices}. Initialization matrix for the three proposed experiment: in (a) for the LM state experiment, in (b) for the M state, and in (c) for the NLM state. The x- and y-axis indicates the measured and prepared state, respectively.}
\end{figure}
\begin{figure}[th!]
	\begin{center}
		\includegraphics[width=1\textwidth]{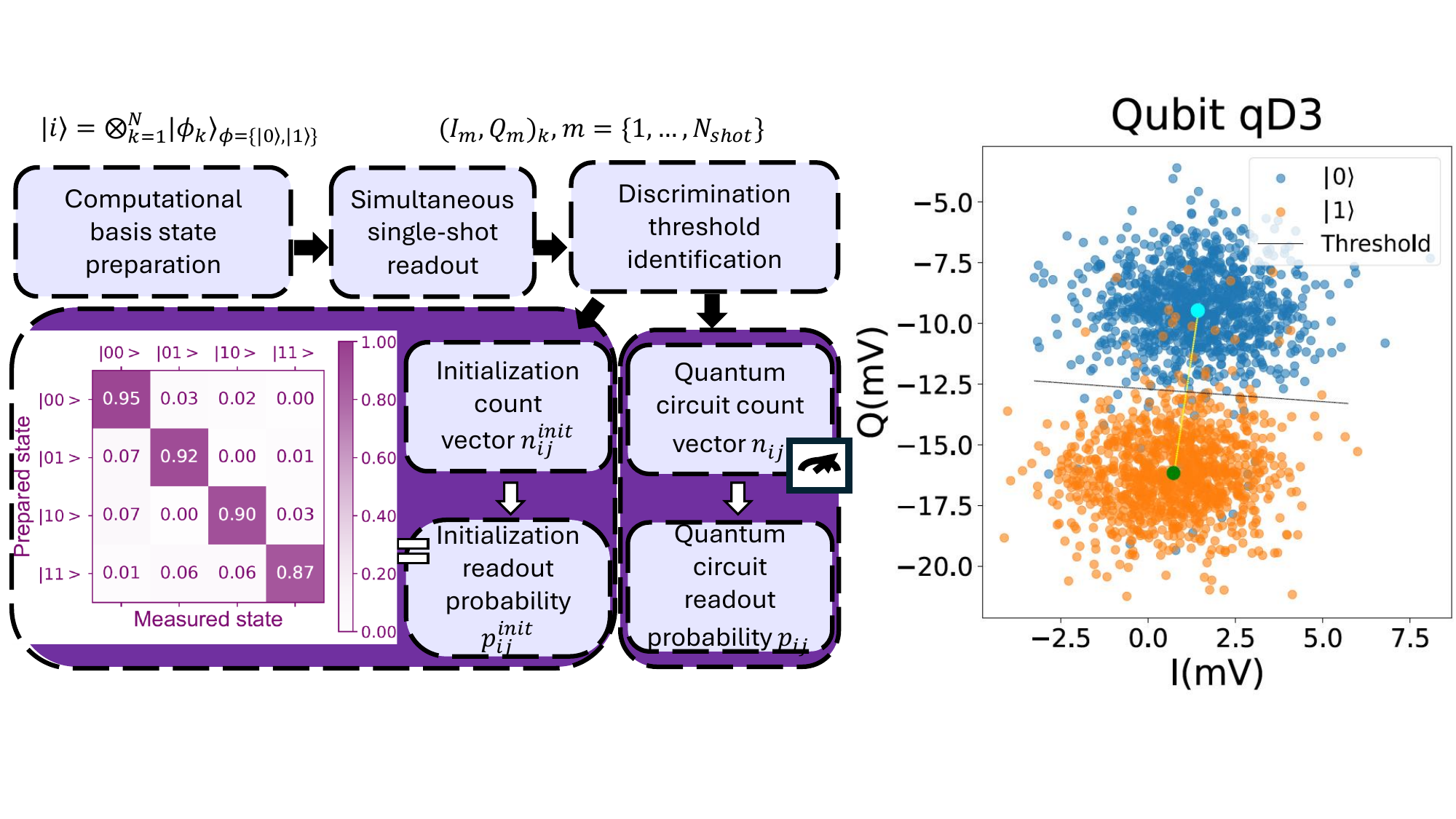}
	\end{center}
	\caption{\textbf{Supplemental Figure - Experimental randomized protocol for magic measurements}. In (a), a schematic representation of the randomized measurement protocol. In (b), the steps involved in the initialization experiment are performed before any quantum circuit. The initialization procedure is used on one hand to calculate the readout fidelity, but most importantly the computational basis states discrimination threshold (black dashed line in the top right inset). This allows to determine the output of any randomized Clifford circuit performed after the preparation of a quantum state.}
	\label{measurement}
\end{figure}

\section{Local magic distillation lemma}
\begin{lemma}[Local magic distillation]
	Given a state $\ket{\psi}\in \hi=\hi_A\ot \hi_B$, an ancillary system in $\ket{0}$, and a Clifford unitary $C$ such that
	\begin{equation}
		C(\ket{\psi}\ot \ket{0})=\ket{\psi'}\ot \ket{\phi}\,,
	\end{equation}
	with $M_2(\ket{\phi})>M^{\rm L}(\psi)$, then $C$ does not allow a decomposition in local unitary operators on $\hi_A\ot \hi_B$, namely it cannot be of the form $C=C_A\otimes C_{BC}$ or $C=C_{B}\otimes C_{AC}$.
	\begin{proof}
		Since $C\in \mathcal{C}$, then $M_2(\ket{\psi'}\ot \ket{\phi})=M_2(\ket{\psi})$. Since $M_2$ is additive, it holds that
		\begin{equation}
			\begin{split}
				&M_2(\ket{\psi'})+M_2(\ket\phi)=M_2(\ket{\psi})\\&=M_2^{\rm NL}(\ket{\psi})+M^{\rm L}(\ket{\psi})
				\implies M_2\left(\ket{\psi'}\right)\\&=M_2^{\rm NL}(\ket{\psi})+M^{\rm L}(\ket{\psi})-M_2(\ket\phi)
			\end{split}
		\end{equation}
		by hypothesis, $M_2(\ket\phi)>M^{\rm L}(\ket{\psi})$, hence $M^{\rm L}(\ket{\psi})-M_2(\ket\phi)<0$: thus we have
		\begin{equation}
			M_2(\ket{\psi'})<M_2^{\rm NL}(\ket{\psi})=\min_{U_A,U_B}M_2[(U_A\ot U_B)\ket\psi]
		\end{equation}
		proving that $\ket{\psi'}$ cannot be the output of the product of local unitaries on $A$ and $B$. 
	\end{proof}
\end{lemma}

\clearpage

%\bibliography{refs_exp}

\end{document}